\documentclass[usenatbib]{mn2e}
\title[Dark Matter In Disk Galaxies - II]{Dark Matter In Disk Galaxies II: Density Profiles as Constraints on Feedback Scenarios}
\author[Hague, P. R., Wilkinson, M. I.]{
P. R. Hague~\thanks{peter.hague@le.ac.uk} \&
M.I. Wilkinson~\thanks{miw6@le.ac.uk}\\
Department of Physics \& Astronomy, University of Leicester, 
  University Road, Leicester LE1 7RH, UK}
  
\usepackage{graphicx}
\usepackage{subfig}
\usepackage{color}
\usepackage{amssymb,amsmath}

\newcommand{\imgext}{pdf}

\begin{document}

\maketitle
\begin{abstract}
The disparity between the density profiles of galactic dark matter haloes predicted by dark matter only cosmological simulations and those inferred from rotation curve decomposition, the so-called cusp-core problem, suggests that baryonic physics has an impact on dark matter density in the central regions of galaxies. Feedback from black holes, supernovae and massive stars may each play a role by removing matter from the centre of the galaxy on shorter timescales than the dynamical time of the dark matter halo. Our goal in this paper is to determine constraints on such feedback scenarios based on the observed properties of a set of nearby galaxies.

Using a Markov Chain Monte Carlo (MCMC) analysis of galactic rotation curves, via a method developed in a previous paper, we constrain density profiles and an estimated minimum radius for baryon influence, $r_1$, which we couple with a feedback model to give an estimate of the fraction of matter within that radius that must be expelled to produce the presently observed halo profile. We show that in the case of the gas rich dwarf irregular galaxy DDO 154, an outflow from a central source (e.g. a black hole or star forming region) could produce sufficient feedback on the halo without removing the disk gas. 

We examine the rotation curves of 8 galaxies taken from the THINGS data set and determine constraints on the radial density profiles of their dark matter haloes. For some of the galaxies, both cored haloes and cosmological $\rho \propto r^{-1}$ cusps are excluded. These intermediate central slopes require baryonic feedback to be finely tuned. We also find for galaxies which exhibit extended cores in their haloes (e.g. NGC 925), the use of a split power-law halo profile yields models without the unphysical, sharp features seen in models based on the Einasto profile.
\end{abstract}


\section{Introduction}

Disk galaxies are presumed by $\Lambda$CDM cosmology to be dominated by dark matter e.g. \citep[e.g.][]{bosma1978}. An understanding of the arrangement of dark matter is therefore necessary for understanding the kinematics and dynamics of these galaxies. Analysis of galactic rotation curves, as well as of N-body cosmological simulations, has produced numerous models describing how dark matter density varies with distance from the centre of a galaxy.

Dark matter only cosmological simulations were found by \cite{dubinski1991} and \cite{navarro1996} to produce haloes with an approximately universal density profile, with density proportional to $r^{-1}$ towards $r=0$ and $r^{-3}$ towards $r=\infty$. These haloes follow a set of scaling relations with virial mass such that they behave as a single parameter family of models \citep{bullock2001}.




The halo density profiles inferred from observing the rotation of disk galaxies appear to contradict this picture. For example, \cite{gentile2004} inferred the density profiles of 5 spiral galaxies and found them to be consistent with flat cores. Rotation curves of 17 galaxies analysed by \cite{bosma2003} were also found to contradict simulations and exclude $r^{-1}$ cusps. This has become known as the cusp-core problem. The possibility of the difference being due to erroneous inference of cores from observations has been discussed, and refuted, in \cite{deblok2002}. This, coupled with an increase in resolution of kinematic data, has at this point resolved such concerns. 

High quality data from The HI Nearby Galaxy Survey, THINGS ~\citep{THINGS} has allowed the generation of more detailed rotation curves, that extend to smaller radii than those used in previous treatments of the cusp/core problem ~\citep{deblok2008}. An analysis of these rotation curves by ~\cite{chemin2011} showed that an Einasto profile $\rho \propto \exp(-r^{\rm n})$, provides a better formal fit than cored profiles such as the Burkert profile ~\citep{burkert1995} or the NFW profile. Also using THINGS data, \cite{oh2011} claimed that a selection of dwarf galaxies (including DDO 154, which we studied using an MCMC method in \cite{hw2013}, hereafter HW13) exhibit $r^{-0.29}$ central slopes. They found all the dwarf galaxies in their study to be inconsistent with $r^{-1}$ inner haloes. 

Simulations have been used to try and resolve the cusp-core problem. \cite{read2005} demonstrated that the baryon physics left out of pure N-body simulations can account for this disparity, in the case of dwarf galaxies, through time asymmetric mass loss (e.g. baryon infall and outflow), and  \cite{governato2010} used supernovae feedback to explain both the flattening of the inner dark matter density profile and the absence of bulges in dwarf galaxies. 

In this paper we attempt to provide an improved modelling of dark matter density profiles in a selection of THINGS galaxies using a more general parameterised density profile, the $\alpha-\beta-\gamma$ profile ~\citep{zhao1996}, and a Markov Chain Monte Carlo (MCMC) method to explore the parameterisation. We also explore the implications of our improved density profiles for out understanding of feedback processes. The structure of this paper is as follows; Section \ref{methodsection} summarises the rotation curve decomposition and MCMC techniques used to derive dark matter density profiles. Section \ref{analysissection} describes a simple analytic model that can be used to constrain formation scenarios for an individual galaxy given its halo density profile and rotation curve. Section \ref{resultssection} discusses the results for individual galaxies, and Section \ref{discussionsection} draws conclusions from the analysis when applied to our full sample of galaxies. 
 
\section{Method}
\label{methodsection} 
 
The rotation curve analysis method used in this paper is described fully in HW13. Here we will briefly summarise our data processing chain: first the method used by \cite{deblok2008} to derive rotation curves from the THINGS velocity fields, secondly the halo density profile we fit to this data and the parameter space defined by it, and then finally the Markov Chain Monte Carlo (MCMC) method that we use to constrain the density profiles of the dark matter haloes. The output of this method is a distribution of halo models, and thus provides not only a best fit model but also robust errors.

\subsection{Baryonic Mass Modelling}

The rotation curve decompositions we used were provided by de Blok, and we describe their generation here for completeness. A more thorough description can be found in \cite{deblok2008}. Data from the Spitzer Infrared Nearby Galaxy Survey~\citep[SINGS: see][]{kennicutt2003} provides surface brightness data that, when mapped onto tilted rings and combined with an estimate of the mass-to-light ratio $\Upsilon$ for the stellar population, can give a radial mass distribution for the stellar component of the galaxy. When modelling 1D rotation curves, we assume it is sufficient to model this as an axisymmetric disk with exponentially decreasing density as a function of radius. 

In \cite{deblok2008} there are two values for $\Upsilon$ used; one derived from a Kroupa IMF and one derived from a version of the Salpeter IMF with the mass reduced by 30\% that is referred to as a diet Salpeter IMF. \cite{bell2001} found that reducing the number of low mass stars relative to the original Salpeter IMF was required in order to satisfy maximum disk requirements in a set of six galaxy evolution models.

Gas can be modelled in the same way, from the THINGS data, but there are fewer issues determining the mass of gas present. HI emission is proportional to the neutral hydrogen present, and a scaling factor of 1.4 is applied that takes into account the amount of helium and metals that will also be present. Again, an exponential disk model is used in order to generate an axisymmetric potential and thus a 1D rotation curve contribution.

Our models do not include molecular gas, but following the method of \cite{deblok2008} we assume that the density distribution of this gas follows that of the stars and is a small fraction of the stellar surface density. There would be no benefit to introducing a parameter describing the ratio of stars to molecular gas, as it would be entirely degenerate with $\Upsilon$, which in all cases besides DDO 154 (due to its very low surface brightness) already encompasses baryonic mass scalings up to, and including, a maximal disk.

There is some uncertainty in the inclinations of the galaxies studied here, which would manifest itself in a rotation curve as a systematic shift in the circular velocity. It would be possible to include the inclination as a parameter in our study, however it would also be degenerate with $\Upsilon$; stellar and gas mass modelling are insensitive to inclination unless significant extinction occurs. This can only happen when the galaxy is near to edge on, and the galaxies in our set have already been selected to have intermediate inclination \citep[see ][]{deblok2008}.  

\subsection{Dark Halo Profiles}

We make three assumptions about the dark matter haloes of the galaxies analysed in this paper

1. They are spherically symmetric

2. The density monotonically decreases with radius

3. The log slope is continuous and differentiable with respect to radius

Satisfying these constraints, a general, spherically symmetric, halo density profile has been selected, which either analytically encloses, or closely emulates, all commonly used density profiles:

\begin{equation}
\rho(x)= \frac{ {\widetilde\Sigma}_{\rm max}}{G} \frac{v_{\rm max}^2}{x^\gamma (1+x^{1/\alpha})^{\alpha ( \beta - \gamma)}} 
\label{abgeq}
\end{equation}

where $x=r/r_{\rm s}$, with $r_{\rm s}$ the scale radius, $v_{\rm max}$ is the peak velocity of the dark matter rotation curve and ${\widetilde \Sigma}_{\rm max}$ is given by

\begin{equation}
{\widetilde \Sigma}_{\rm max} = { \rho_{\rm s} r_{\rm max} \over M(r_{\rm max})}
\end{equation}

where $r_{\rm max}$ is the radius at which the profile's rotation curve reaches its maximum velocity and $M(r_{\rm max})$ is the mass enclosed at that radius. A derivation and more complete description of this can be found in HW13.

This is a parameter transform of the well known $\alpha-\beta-\gamma$ halo \citep{zhao1996}

\begin{equation}
\rho (x) = {{\rho_s} \over {x^\gamma (1+x^{1/\alpha})^{\alpha(\beta-\gamma)}}}
\end{equation}

where $\rho_{\rm s}$ is the scale density.

Because double power laws with respect to radius can be approximately modelled (provided the transition between the two powers is smooth) by a single power law over a finite range, we assert that our two power model can represent any profile containing more power laws, over a finite radial range, as long as a similar condition of smoothness is met.


\subsection{Markov Chain Monte Carlo}

A Markov Chain Monte Carlo (MCMC) method \citep{hastings1970} is used to integrate over the parameter space defined by the profile given in equation \ref{abgeq} along with two disk scaling parameters; a mass-to-light ratio multiplier $f_\Upsilon$ which scales the magnitude of the stellar contribution \citep[relative to the mass-to-light ratio $\Upsilon$ for the diet Salpeter IMF case in ][]{deblok2008}. We consider $f_\Upsilon$ in the range [0.316, 3.16] which encloses the Kroupa IMF and the factor of 2 variation in infrared IMF suggested by \cite{bell2001}. 
We use {\tt CosmoMC} \citep{lewis2002}, in a generic mode, together with a likelihood function of our own as our MCMC engine. MCMC is a Bayesian method that generates a probability distribution in the parameter space of a set of models. Bayes' Theorem states that 

\begin{equation}
P(M[{\bf x}] | D) = {P(D | M[{\bf x}]) P(M[{\bf x}]) \over P(D)}
\end{equation}

where $M$ is a proposed model defined by parameter vector ${\bf x}$, and $D$ is the data. Assuming that $P(D)=1$, we can then generate a series of values for $P(D | M)$ via a Markov chain process. Starting at a random point, the algorithm generates a new candidate point based on a random step drawn from a probability distribution, which in this case we take to be Gaussian. The new point ${\bf x}^\prime$ is added to the Markov chain with a probability

\begin{equation}
P_{\rm accept} = {\rm min} \left[ {P(D | M[{\bf x}^\prime]) \over P(D | M[{\bf x}])} , 1 \right]
\label{mcmceq}
\end{equation}

if the new point is not accepted, the current point is added again to the Markov chain. This process is repeated and, provided the chain is able to converge, produces a distribution of $P(D | M[{\bf x}])$, which can be combined with the prior distribution $P(M[{\bf x}])$ to calculate the probability distribution we seek. 

A condition required for an MCMC run to converge well is that the step size (measured in our Gaussian case by a standard deviation $\sigma$ in each parameter) is appropriate to the distribution. This can either be done by trial and error, or by a step size that adapts during the course of the run. We opt for the second approach, updating the probability distribution of candidate points relative to the current point based on a covariance matrix generated from the most recent half of the models generated. {\tt CosmoMC} then samples $P^{1/T}(D | M[{\bf x}])$ where $T$ is a temperature value which we set equal to 1 (reducing to Equation \ref{mcmceq}) unless there are specific problems finding a constraint, where we use $T=2$.

A further condition for convergence is that the parameter space be able to cover the entire range of models that are likely to fit the data. In our case, this condition is of most concern for those haloes where $\gamma \rightarrow 0$, which are required to model the flat-cored \cite{burkert1995} halo. Boundary effects at the edge of the parameter space not only cause problems integrating there due to cutting off part of the Gaussian selection function, but also necessarily mean that the distribution is not normalised. Strictly, MCMC produces a non-normalised distribution, but the distribution can be normalised correctly if the probability tends towards zero at the boundaries of the parameter space, and the distribution can be assumed to have no models with nonzero probability outside this space. To resolve this boundary issue, we mirror the parameter space around $\gamma=0$. We allow values to become negative, but take the absolute value as an input to the density profile. 

We use 8 independent MCMC chains, each using approximately $5\times10^6$ models (the code terminates once the first chain completes, but as the run time for the chains is quite consistent this does not usually cause the chains to be unduly shortened.) The first $1\times10^5$ models in each chain are treated as burn in, the process by which the MCMC algorithm initially finds and moves into the main body of the distribution, and discarded.

Previous work on rotation curves such as \cite{chemin2011} and \cite{oh2011} has tested a small number of proposed density profiles against the data; these profiles represent single points (or in the case of the Einasto profile, curves) within the parameter space we explore. By fully characterising the parameter space, we are able to put these previous fits in a wider context.

\section{Analysis} 
\label{analysissection}

Once we have generated distributions of density profiles for each galaxy, we investigate their systematic properties, with reference to previous claims regarding the nature of dark matter haloes. Our primary focus is to find properties of the dark matter haloes that could constrain formation scenarios. 

In previous work \citep[see e.g.][]{vandenbosch2001}, haloes whose density profiles became flat as $r \rightarrow 0$ are referred to as 'cored' whereas haloes that show density profiles approaching $\rho \propto r^{\rm -n}$ where $n \geq 1$ are referred to as cusped. Terminology for intermediate haloes (i.e. those with $0 < n < 1$) is unclear. For the purposes of our discussion, we consider such haloes neither fully cusped nor fully cored.

The parameter $\gamma$, the asymptotic log slope of the profile as $r \rightarrow 0$, is a property of the profile that describes the shape of the halo outside the data range. Therefore we generate, from the halo profiles, an alternative parameter $\gamma_{\rm in}$. This is the log slope of the dark matter halo density profile at the innermost data point of the measured rotation curve. As the data sets all start at a different radius, direct comparisons between galaxies using $\gamma_{\rm in}$ are not possible. It is, however, a useful quantity because if it is substantially above zero, it excludes the existence of a core within the data range. 

Another value we extract from our density profiles is $r_1$. This is the radius at which the log slope (which must monotonically decrease for all profiles) reaches $-1$. As we will show, this a well constrained and useful scale radius\footnote{In some work \citep[e.g. ][] {chemin2011}, the radius $r_2$ is also used (sometimes written as $r_{-2}$, but we exclude the sign in order to maintain consistency with the positive sign of the parameters in the $\alpha-\beta-\gamma$ profile.) For individual profiles (such as Burkert profiles) the relationship between these is a straightforward, constant ratio.}. In order to demonstrate its utility, we construct a simple model for the formation of the galaxies we study here.

If we assume that the starting point for a dark matter halo is an NFW profile (or some other cosmological profile that tends towards a log slope of -1 at $r=0$), then the density profile must have a logarithmic slope of $d {\rm log} \rho / d {\rm log} r = -1$ or steeper at all radii, so any part of the density profile that has a shallower slope can clearly be stated to be inconsistent with a cosmological profile derived from pure dark matter N-body simulations. Thus $r_1$ gives us a radius to which the effects of any process which modifies the shape of the halo must reach in order to account for the current density profile, and we now show that this is a conservative estimate.

The choice of the parameter $r_1$ is based on the minimisation of assumptions. To construct an exact model of an observed galaxy at the time of its formation would require knowledge of the density of the Universe and assert the exact correctness of a particular halo model such as NFW or Einasto (whereas we only assume that the halo had a cusp of some kind). Using $r_1$ allows a degree of physical modelling without having to resolve these unknowns.

In order to investigate the origins of non-cosmological (i.e. $\gamma<1$) haloes, we construct a simple model of feedback, where outflows of gas on timescales shorter than the dynamical time of the halo can impart energy to the halo and alter its density profile. The source of this feedback can either be an accreting central black hole, supernovae, or other stellar feedback.

Our model features an inhomogenous spheroid of gas in a spheroid of dark matter, and a disk. Gas is removed entirely from the galaxy instantaneously and permanently during an outflow \citep[realistic, given velocities of various possible outflows in e.g.][]{governato2010,king2003}. Spheroidal symmetry allows us to make a number of simplifying assumptions in the discussion which follows in both the spherical case, and the oblate spheroidal case \citep{BT}. Hereafter we only discuss radius, however the arguments still apply using the equivalent coordinate in an oblate spheroidal system.

Given the assumed geometry, dark matter particles can only respond to changes in the mass interior to their current position. If a certain portion of this mass is removed, a particle will respond by moving outwards. Assuming a much longer timescale for the removed mass to fall back in (if it does at all) the upper limit to the instantaneous change in the total energy of dark matter particles at any particular radius, taking the potential to be relative to $r_1$ is the change in potential interior to that radius. As baryons and dark matter move in the same potential, we assume a mass of baryons outflowing past $r_1$ can bring the same mass of dark matter with it. The change in dark matter mass of the halo interior to $r_1$ is obtained from

\begin{equation}
M_{\rm I, dark} (r_1) = M_{\rm F, dark}(r_1) (1 - f_{\rm b})^{-1}
\label{masseq}
\end{equation}

where $M_{\rm I,dark}(r)$ is the initial mass interior to $r$ and $f_{\rm b}$ is the fraction of mass that is in baryons and thus available for an outflow.  We assume that the available gas is arranged in a diffuse spheroid the follows the density profile of the dark matter. In this case we are modelling a single outflow, or multiple outflows where any gas falling back in negates the effects of its initial outward movement exactly. In reality, infall happens over a much larger timescale than outflow \citep[see e.g.][]{governato2010}, but modelling this is beyond the scope of the simple model we are constructing here, and will be addressed in later work.

We assume here that the amount of dark matter removed from inside a radius $r$ is approximately equal to the amount of gas removed. To verify this, we ran a series of $2 \times 10^6$ particle N-body simulations using {\tt falcON} \citep{dehnen2000,dehnen2002}, using spherical initial haloes with inner log slopes $\gamma=0, 0.25, 0.5, 7/9, 1$, and removing a fraction of mass equal to $f_b=0.02, 0.04, 0.08, 0.16, 0.24, 0.48, 0.64, 0.90$. Mass is removed equally from all of the particles in the simulation. Under conditions of spherical symmetry this does not produce a different result from removing all the mass interior to $r_1$, and does not require an extra step of calculating the position of $r_1$. 

\begin{figure}
  \includegraphics[width=\linewidth]{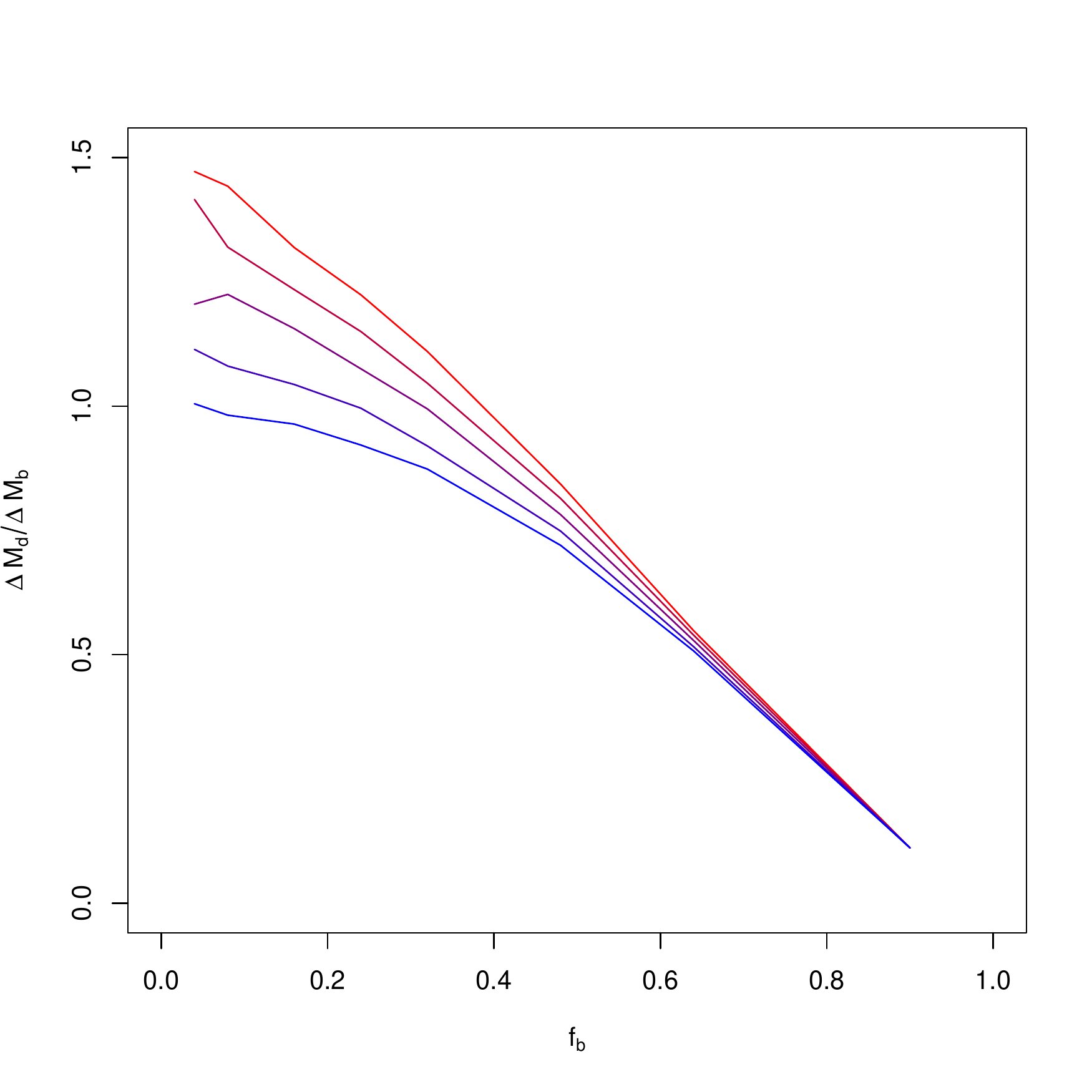}
  \caption{Ratio of expelled dark matter $\Delta M_{\rm d}$ to expelled baryons $\Delta M_{\rm b}$, from N-body simulation, for a range of initial conditions. Red to blue denotes increasing log slope (0, 0.25, 0.5, 7/9, 1). Simulations were run for baryon fractions $f_b=0.02, 0.04, 0.08, 0.16, 0.24, 0.48, 0.64, 0.90$. Deviations of the curves at low $f_b$ are due to $\Delta M_{\rm b}$ being a smaller multiple of $M_{\rm particle}$}
  \label{efficiency}
\end{figure}

We compared mass before and after, within the scale radius, and found that for a cosmological baryon fraction, the fraction of dark matter carried away with gas outflow is of order unity, as shown in Figure \ref{efficiency}.

The starting conditions for these simulations were constructed in equilibrium using {\tt mkhalo} from the same software suite - in reality the response of the dark matter to gas outflow is dependent on the velocity structure of the halo. This velocity structure depends on the precise history of the halo, as shown in \cite{read2005} where halo cores are formed by moving particles from tangential to radial orbits. Thus the exact order of the curves in Figure \ref{efficiency} should not be taken as strictly realistic, but it does demonstrate that the ratio between the dark matter removed and the gas removed at a certain radius should be of order unity. 

From equation \ref{masseq}, assuming an initial and a final mass profile, we can compute the spheroidal baryon content at the time the outflow(s) occurred. In reality this could be raised by contraction of the gas content of the galaxy, or lowered due to baryons leaving the spheroidal gas component through either clumping, or becoming part of the disk, and either way being too dense to participate in an outflow event. However, we are able to produce a consistent model to predict $f_{\rm g}$ - the fraction of matter available as gas of sufficiently low density to be part of the feedback process.

Assuming the dark matter halo mass interior to the outermost data point we have observed is unchanged through feedback, there exists an NFW profile (using the scaling relation established in cosmological simulations) that gives a value for $M_{\rm I}(r_1)$ and the model above then gives $f_{\rm g}$. In Section \ref{discussionsection} we determine this value for this set of galaxies, and discuss whether they are realistic.


\section{Results}
\label{resultssection}

Table \ref{galaxytable} shows relevant physical information for the galaxies in our sample and some detail of the MCMC output. We have sampled $~4\times10^7$ models in each case, using 8 independent MCMC chains which have converged to similar distributions. The quantitative convergence statistic we use here, provided by the {\tt getdist} program which accompanies {\tt CosmoMC}, computes the variance of the means of the chains $\sigma(\hat{x})$, divided by the mean of the variances of the chains $\hat{\sigma}(x)$, and must be smaller than unity for a set of chains to be converged. This statistic is calculated for each parameter, and the largest value is taken to be an overall measure of convergence. 

\begin{table*}
\begin{tabular}{lccccccl}
Galaxy & $V_{\rm c, max} ({\rm kms}^{-1})$ & $A_{\rm r, 1kpc} ({\rm kms}^{-1}) $ & Absolute $B$ Magnitude & HII/HI ratio & Models & T & $\sigma(\hat{x})/\hat{\sigma}(x)$\\
& (1) & (2) & (3)  & (4) & (5) & (6) & (7) \\ \hline
DDO 154 & $36.8\pm13.3$ &  $1.43\pm_{0.53}^{0.14}$ & $-14.23$  & $  $ & 38880635 & 1 & 0.0085 \\ 
NGC 2976 & $51.1\pm20.85$ &  $2.18\pm_{0.55}^{0.77}$ & $-17.78$ & $0.36$ & 39310884 & 1 & 0.0098 \\ 
NGC 7793 & $80.9\pm27.6$ & $3.41\pm_{0.48}^{0.64}$ & $-18.79$ & $  $ & 39539819 & 1 & 0.0529 \\ 
NGC 2403 & $125.9\pm18.95$ & $2.60\pm_{0.48}^{0.59}$ & $-19.43$ & $  $ & 34880553 & 2 & 0.1315 \\ 
NGC 925 & $75.7\pm29.95$ & $9.45\pm_{2.98}^{0.64}$ & $-20.04$ & $0.04$ & 38180473 & 1 & 0.0959 \\ 
NGC 3621 & $136.8\pm20.6$ & $5.52\pm_{3.21}^{0.94}$ & $-20.05$ & $  $ & 38025138 & 1 & 0.3916 \\ 
NGC 3198 & $145.6\pm18.25$ & $1.50$ & $-20.75$ & $0.04$ & 36688012 & 1 & 0.3113 \\ 
NGC 3521 & $201.9\pm63.35$ & $3.12\pm_{1.77}^{12.67}$ & $-20.94$ & $0.38$ & 36845437 & 2 & 0.7693 \\ 
\hline
\end{tabular}
\caption{Galaxies studied in this paper, ordered by absolute $B$ magnitude. (1) Maximum circular velocities we obtained by reduced $\chi^2$ fitting of an $\alpha-\beta-\gamma$ profile to the rotation curve data. (2) Amplitudes of non-circular motions, in the inner 1 kpc of each galaxy, from \protect\cite{trachternach2008}. NGC 3198 only has one data point inside this radius so no bounds are available. (3) Absolute $B$ magnitude from \protect\cite{THINGS}. (4) Molecular gas to atomic gas ratio, calculated from HERACLES data \protect\citep{leroy2009}, where available. (5) The number of models sampled, excluding a burn in of 10,000 models in each of 8 chains. (6) The temperature parameter used with {\tt CosmoMC}, which samples $P^{1/T}$ of each model. (7) Largest (i.e. worst) value of the statistic used to check that chains are converged on to the same likelihood distribution. See text for details.}
\label{galaxytable}
\end{table*}

The galaxies we study are modelled by \cite{deblok2008}, utilising HI velocity maps from THINGS to derive the rotation speed, the same HI data to derive a radial gas density profile, and $3.6{\rm \mu m}$ maps from SINGS (along with assumed $\Upsilon$ ratios) to derive the stellar components. Some of the galaxies in this sample are modelled with a single stellar disk, whilst some are modelled with a stellar disk and a separate bulge component. 

\begin{figure*}
  \includegraphics[width=\linewidth]{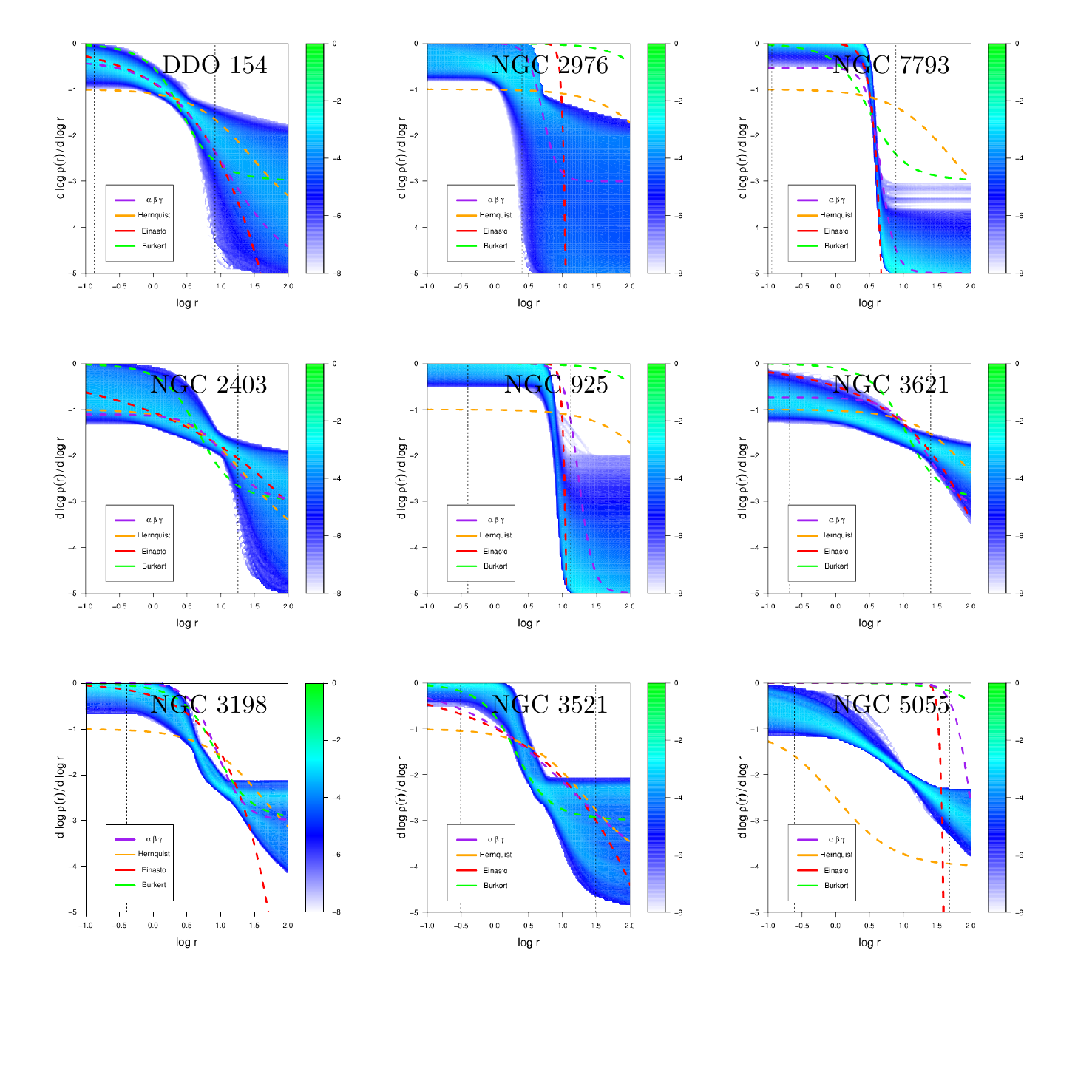}
  \caption{Overlay of all models produced by the MCMC algorithm for each galaxy, in ${\rm log} r - {\rm d log} \rho / {\rm d log} r$ space. The log density scale on the right of each plot is relative to total number of models. Vertical dotted lines indicate the radial range of the observed rotation curve data, with the inner limits for NGC 2976 and NGC 2403 being below $log r = -1.0$. Dashed curves are least squares minimisation fits of common density profiles (Green: Burkert profile. Orange: Hernquist profile. Purple: $\alpha-\beta-\gamma$ profile. Red: Einasto profile.) NGC 5055 is included for comparison although it is one of the galaxies rejected from our analysis. See text for a detailed discussion.}
  \label{dcmpgrid}
\end{figure*}

\begin{figure*}
  \includegraphics[width=\linewidth]{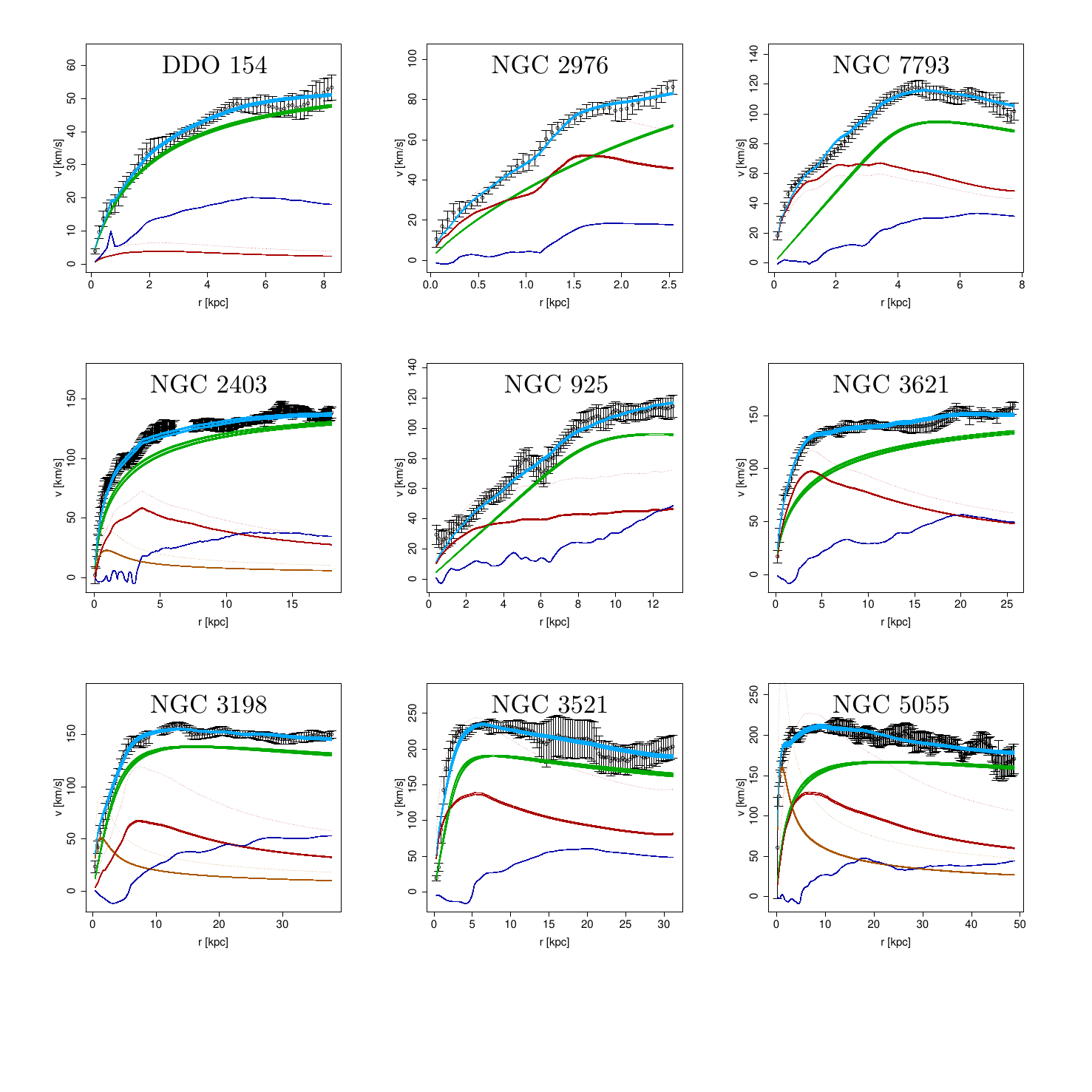}
  \caption{Rotation curve decompositions showing the dark matter haloes corresponding to the most populated bin in the parameter space distribution. Black points represent the observed HI circular velocities, the dark blue curves are the inferred contribution of neutral atomic gas, the solid red curve is the stellar contribution in the best fitting bin, the dotted red curve is the input value for the stellar contribution (which assumes a diet Salpeter IMF), the green curve is the dark matter contribution and the light blue curve is the resulting combined curve. The green and light blue curves have a thickness as they are the superposition of multiple models in the most populated bin. Two stellar curves are present in NGC 2403, NGC 3198 and NGC 5055 as these galaxies were modelled with two stellar components in \protect\cite{deblok2008}}
  \label{curvegrid}
\end{figure*}

Figure \ref{dcmpgrid} shows the distribution of halo log slope profiles for all accepted models in the MCMC chains. We categorise as having cored haloes NGC 925, NGC 3198, NGC 3521, and (with a weaker constraint) NGC 2976. We find cusps in NGC 2403 and NGC 3621, and intermediate inner log slopes in DDO 154, NGC 3198 and NGC 7793. Figure \ref{curvegrid} shows the rotation curve decomposition of the most populated bin of parameter space in the MCMC chains.

\subsection{DDO 154}


This is the least massive galaxy in the set we are examining, having a mass of $3\times10^9M_\odot$ according to \cite{carignan1998}, and thus has lower rotation speeds overall than the other galaxies. Given that its low surface brightness makes the dark matter halo parameters easier to constrain, we used it as a test case for the method that we discussed in HW13. That paper contains some analysis of the result.

The HI velocity field of this galaxy is asymmetric in the outer parts \citep[illustrated in Figure 81 in][]{deblok2008}, but as we are concerned mainly with the potential at small r, this is not an obstacle to the constraint of $\gamma_{\rm in}$. The rotation curve is well constrained and, based on the errors which are calculated from the difference between the two sides of the rotation curve, quite symmetrical in the region of interest. We find a well constrained $\gamma_{\rm in}$ value, with an intermediate log slope at the innermost datapoint ($r=135.7{\rm pc}$.) Unsurprisingly this value is unaffected by the chosen value of $f_\Upsilon$ due to the low contribution of the stellar disk to the rotation curve. 

In HW13, the gas curve for this galaxy was represented by the rotation curve of a smooth exponential disk. We have opted to no longer do this, first because the gas contribution does not significantly alter the result anyway, and secondly because any smoothing that is necessary should be handled by the MCMC process. We now directly use the gas velocity data from \cite{deblok2008}. This reasoning is supported by the fact that we indeed get the same value for $\gamma_{\rm in}$, within error, as that reported in HW13.

\subsection{NGC 2403}



The data for this galaxy have large error bars relative to their scatter \citep[Figure 70, ][]{deblok2008}. However we demonstrated in HW13, using test data, that our method is robust to this issue. 

\begin{figure}
  \includegraphics[width=\linewidth]{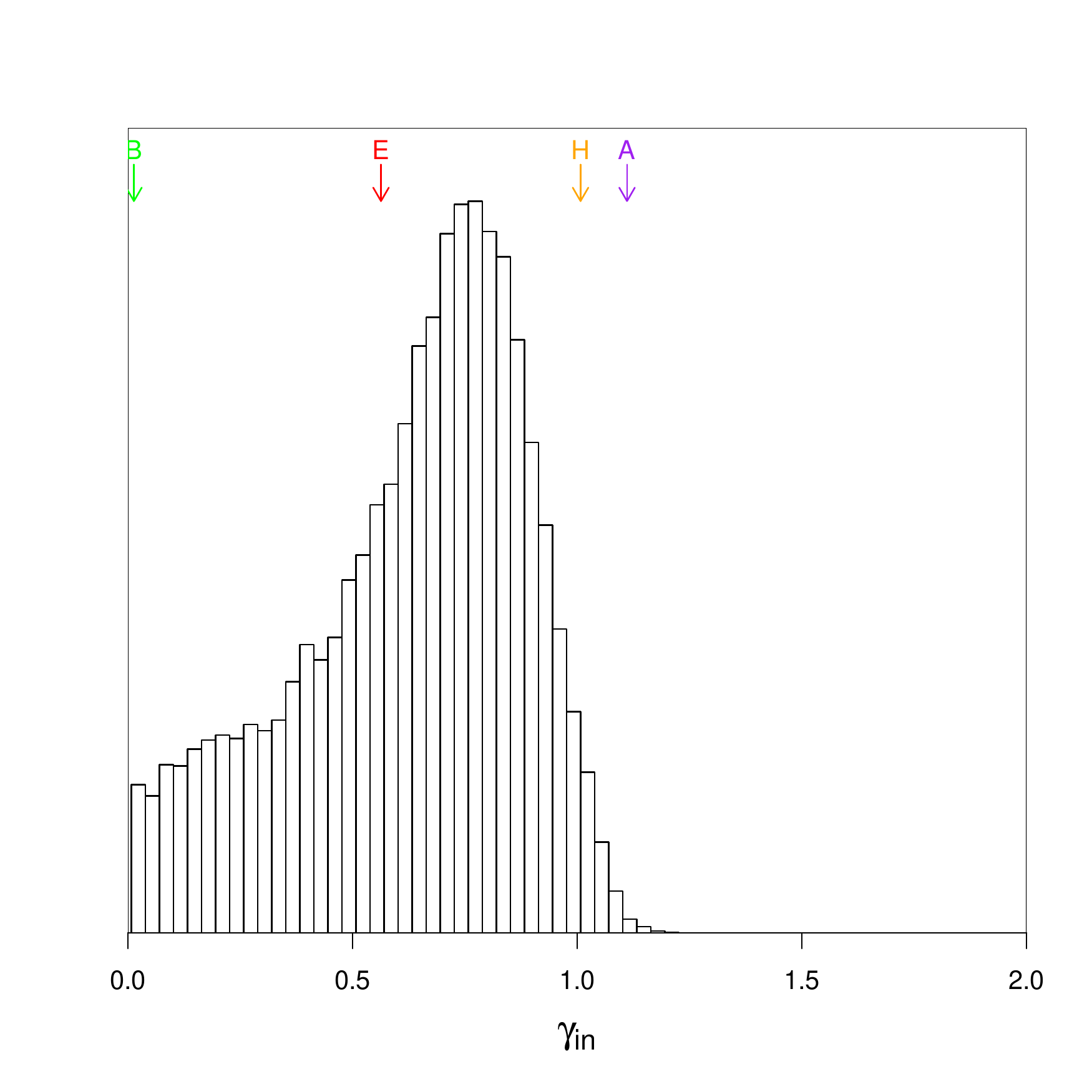}
  \includegraphics[width=\linewidth]{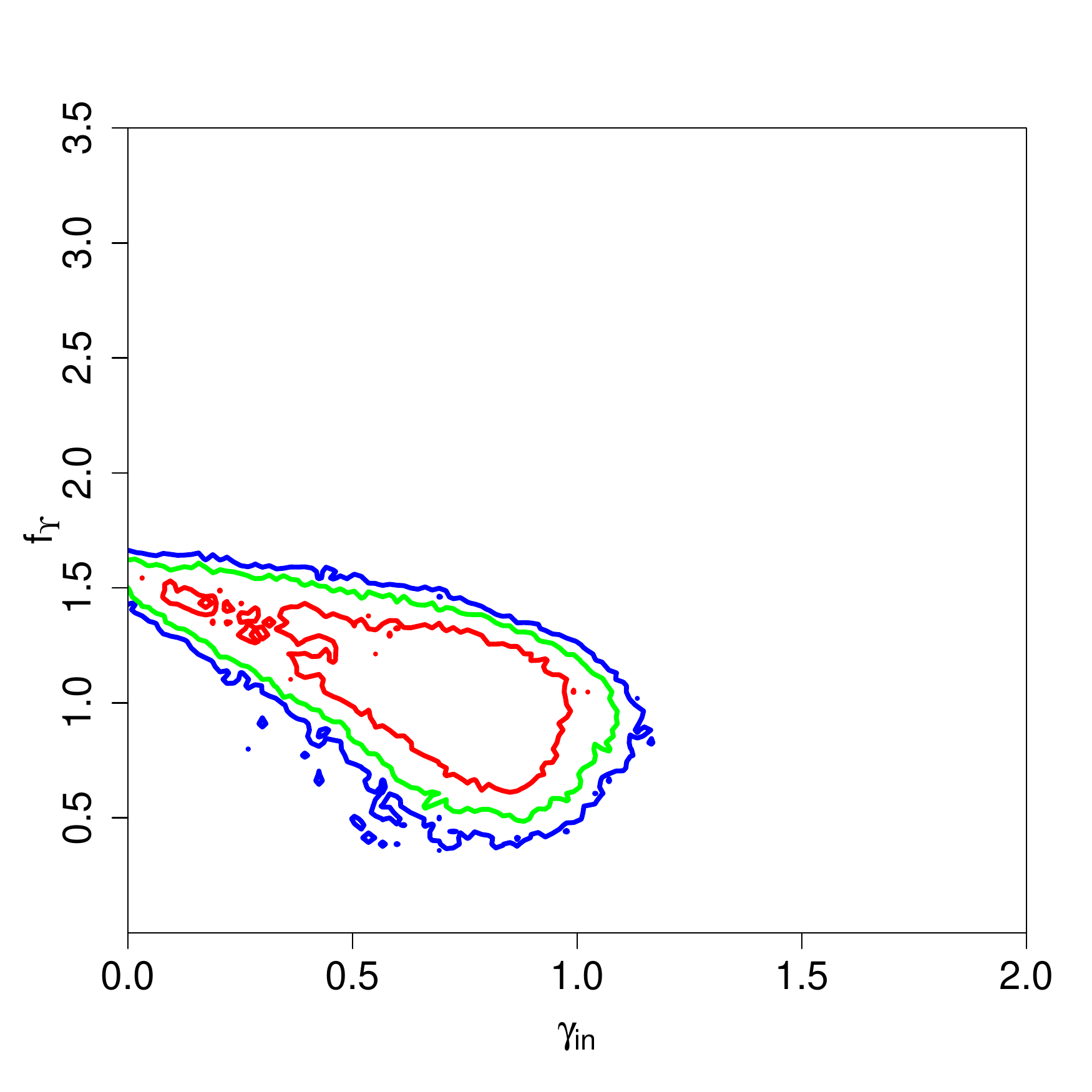}
  \caption{{\bf Top} histogram of $\gamma_{\rm in}$ values produced by the MCMC analysis of NGC 2403 with a freely varying mass-to-light ratio multiplier ($f_\Upsilon$). Arrows indicate the logarithmic slope at this radius of single profile fits (Green: Burkert profile. Orange: Hernquist profile. Purple: $\alpha-\beta-\gamma$ profile. Red: Einasto profile.) {\bf Bottom} Contour plot of $f_\Upsilon$ versus $\gamma_{\rm in}$ demonstrating the nature of the degeneracy between the two parameters. The red contour encloses 0.68 of all models in the MCMC chains, the green contour encloses 0.95 and the blue contour encloses 0.99.}
  \label{NGC2403A}
\end{figure}

\begin{figure}
  \includegraphics[width=\linewidth]{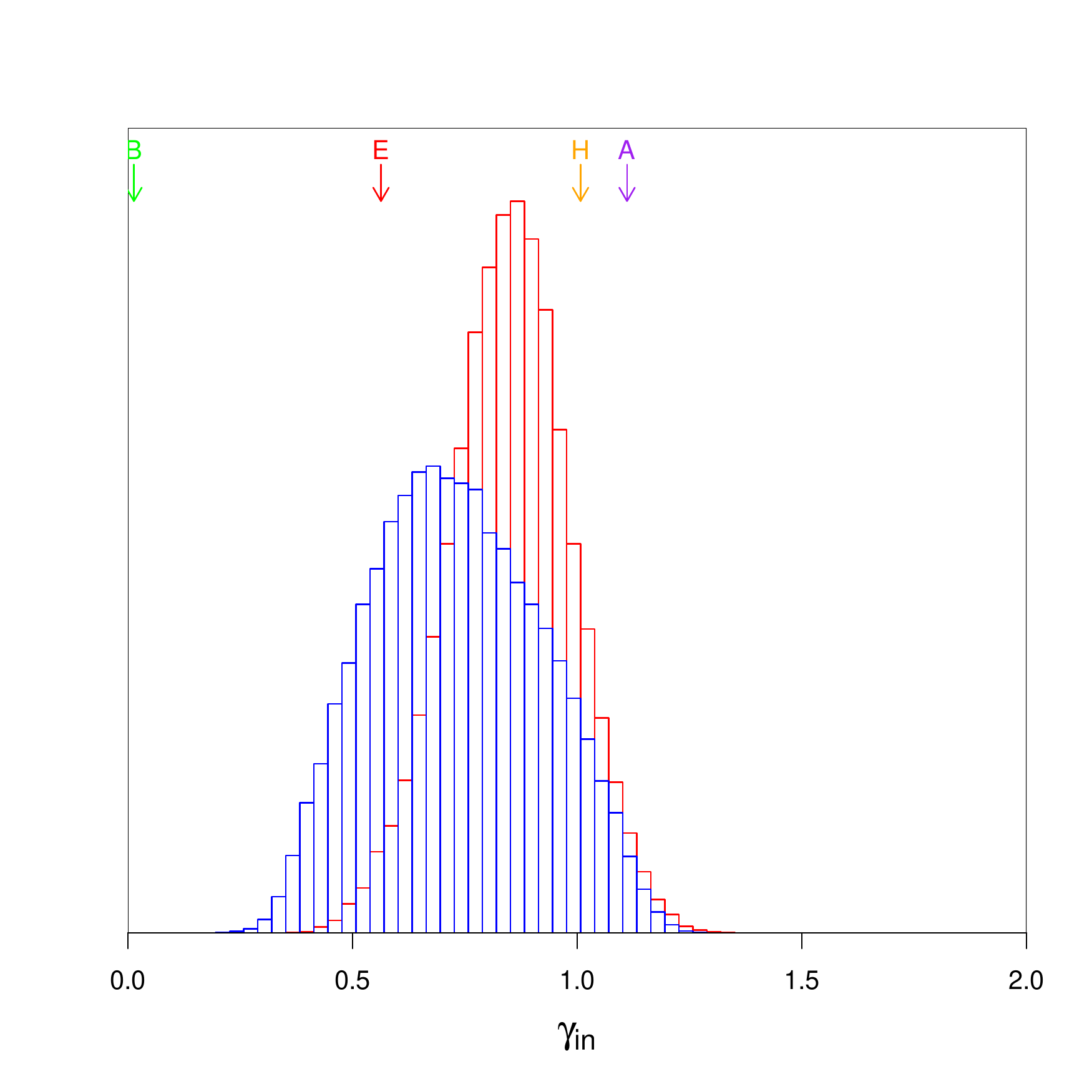}
  \caption{Histograms of $\gamma_{\rm in}$ values produced by MCMC analyses of NGC 2403 with fixed stellar IMF. {\bf Blue} with a fixed $f_\Upsilon$ derived from a diet salpeter IMF. {\bf Red} with $f_\Upsilon$ derived from a Kroupa IMF \protect\citep{kroupa2001}. Arrows are as in Figure \ref{NGC2403A}.}
  \label{NGC2403B}
\end{figure}

\begin{figure}
  \includegraphics[width=\linewidth]{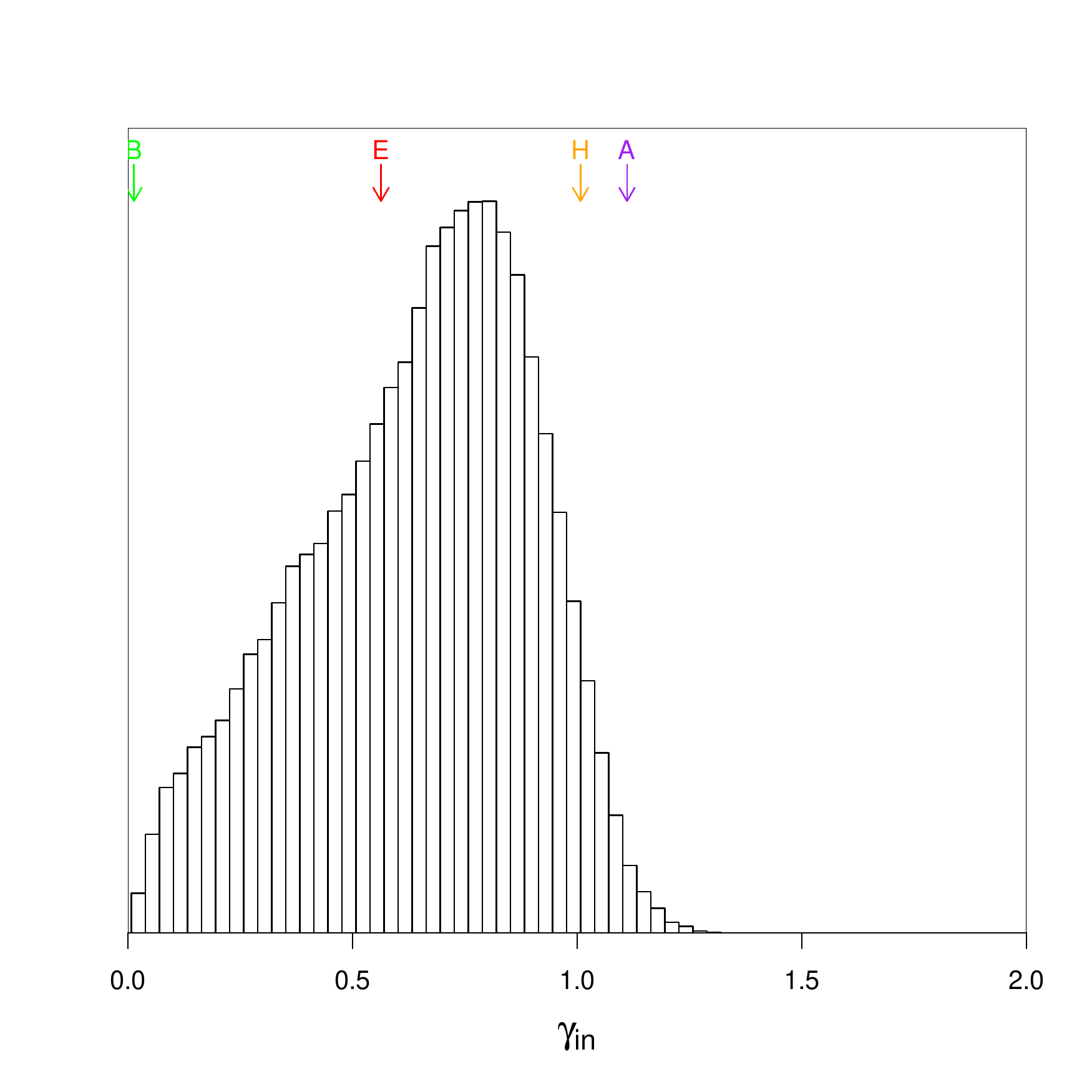}
  \caption{{\bf Top} histogram of $\gamma_{\rm in}$ values produced by the MCMC analysis of NGC 2403 with a freely varying mass-to-light ratio multiplier ($f_\Upsilon$), and a temperature $T=2$. Key as in Figure \ref{NGC2403A}. See text for detailed discussion.}
  \label{NGC2403C}
\end{figure}

The constraint on the inner log slope is shown in Figure \ref{NGC2403A}. With a freely varying mass-to-light ratio, the distribution of $\gamma_{\rm in}$ is clearly not Gaussian, showing a plateau towards more cored values. This part of the parameter space is noisy, and the level of the noise is not consistent across parameter space. Whilst formally converged ($\sigma(\hat{x})/\hat{\sigma}(x)=0.73$), the presence of the plateau reveals something interesting about the parameter space, as well as pointing to the need for judicious use of convergence statistics. The peak at the high end of the $\gamma_{\rm in}$ distribution appears in all chains. Figure \ref{NGC2403B} shows that the inner log slope has a single smooth peak when $f_\Upsilon$ is fixed by assuming either a diet Salpeter or Kroupa IMF, in both cases consistent between chains, indicating that the additional tail in the free $f_\Upsilon$ case is due to a degeneracy (shown in Figure \ref{NGC2403A}) between the dark matter halo parameters and the mass-to-light ratio. This is to be expected for a higher surface brightness galaxy - at higher values of $f_\Upsilon$, the stellar contribution is able to model the rotation curve alone, and the MCMC then finds fits with low dark matter density where the exact shape of the halo is less important. 

The small, cored plateau represents models which attempt to fit maximal disk models to the data, and such models require very high values of $f_\Upsilon$. Furthermore, the fact that these models cannot form a second peak in the distribution indicates they do not reproduce the data as well as those in the peak. We now explore this in more detail. The maximal disk part of parameter space occupies a large volume, due to the fact that the shape parameters of the dark matter halo are no longer significant, and can be freely varied without compromising the quality of the fit. This can bias the MCMC chains towards that volume, so if models there were of a higher fitness than those in the peak, the plateau region would be strongly favoured in all chains. If we disregard these models on this basis, we can conclude that NGC 2403 has a moderate cusp. As shown in Figure \ref{NGC2403B} a cusp is also indicated when using a fixed $f_{\Upsilon}$ derived from either Kroupa or diet Salpeter IMF.

We took a subsample of 1423 models from the MCMC chain, that were randomly selected after burn-in with a probability of $10^{-4}$ for each model, and then divided this into 2 subsamples either side of $\gamma_{\rm in} = 0.5$. We found no difference in reduced $\chi^2$ for either side, to 2 significant figures. Both produced a 90\% confidence interval of $\chi^2_{\rm red} = 0.30, 0.31$. Given the very definite preference of the MCMC chain, and the fact that values of $\chi^2_{\rm red} <  1$ are of little use for comparison, we conclude that the algorithm is still able to produce a meaningful result. We also note that the number of degrees of freedom is treated as a constant when using $\chi^2_{\rm red}$, but the nature of the problem means that in some parts of the parameter space not all of the parameters contribute significantly to the fit - for example, if $v_{\rm max}$ is low and $f_\Upsilon$ is high then the shaping parameters of the dark matter halo can be freely varied whilst maintaining a nearly constant proposed rotation curve. This inherent weakness in $\chi^2_{\rm red}$ does not apply to our MCMC method, as we use the ratio between a proposed model and the current one to determine the next step in the chain, and thus $\chi^2_{\rm red}$ is equivalent to $\chi^2$ (which we use) as the degrees of freedom cancel.

In order to try and produce a better convergence for the tail, we reran the MCMC algorithm using a temperature $T=2$ (and thus sampling $P^{1/2}$ rather than $P$.) The result was more consistent across independent chains, and produced a smoother distribution overall when the chains are summed, as shown in Figure \ref{NGC2403C}. This produces the same constraint on $\gamma_{\rm in}$ within $1\sigma$ errors. The numeric values discussed in later sections are derived from this version. 

\subsection{NGC 2976}

\begin{figure}
  \includegraphics[width=\linewidth]{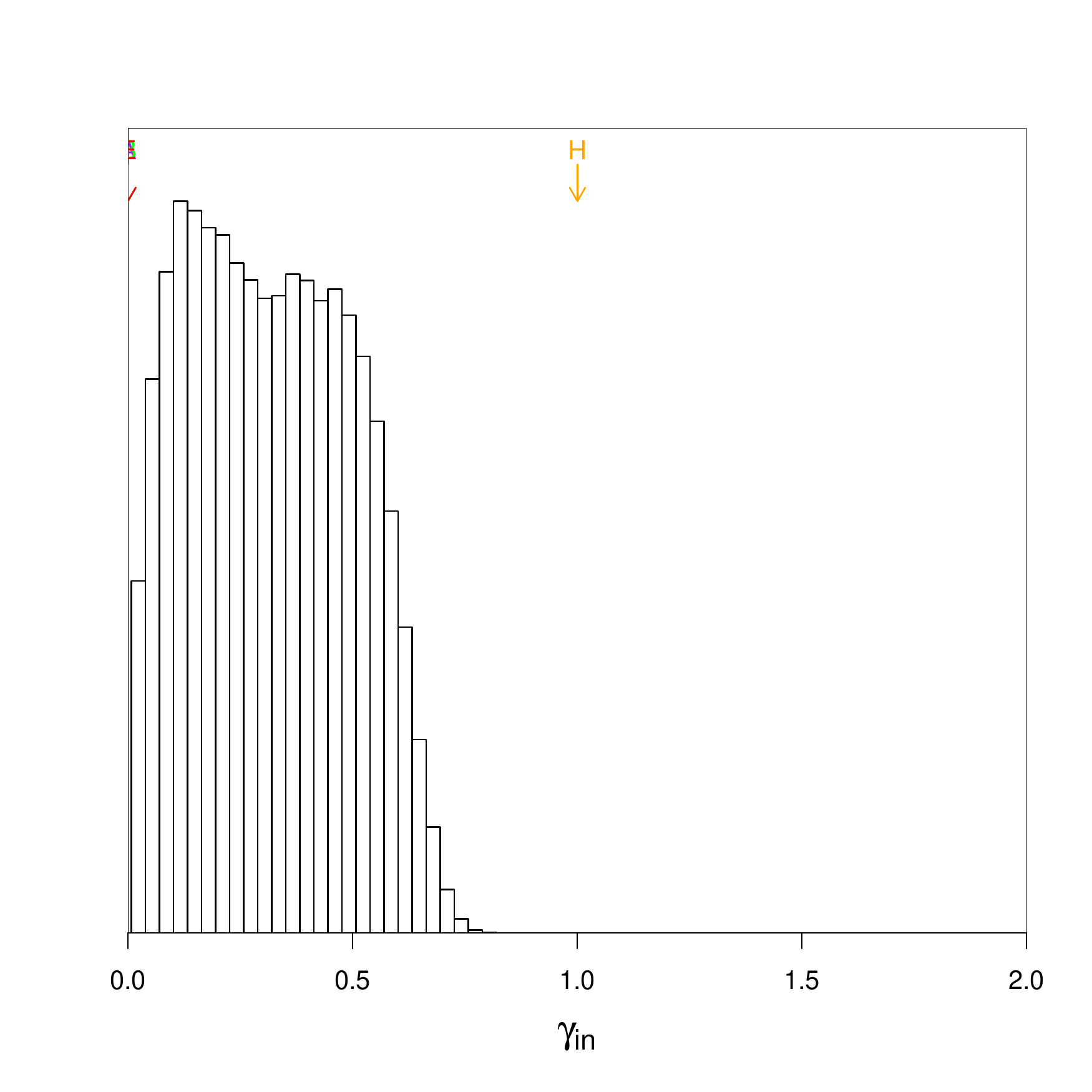}
  \caption{Histogram of the $\gamma_{\rm in}$ value of all the models produced by the NGC 2976 MCMC run. }
  \label{NGC2976gammain}
\end{figure}

This galaxy has a high surface brightness, so is not expected to produce as strong a constraint on the dark matter halo as some of the other galaxies have done. This is borne out by the $\gamma_{\rm in}$ distribution in Figure \ref{NGC2976gammain}.

As shown in Figure \ref{curvegrid}, the rotation curve data for NGC 2976 show a sharp increase just outside 2kpc, and it is possible to get a mathematically credible fit to the data without treating this as a feature at all. However, the shape of the stellar contribution to the rotation curve suggests that this is the region where dark matter should begin to dominate. Interior to this, the rotation curve follows the features of the stellar contribution well. Therefore, an adequate fit (from the perspective of $\chi^2$ only) can be obtained using $f_\Upsilon \geq 1$ and ignoring the last few data points.

We find, however, that MCMC selects lower values of $f_\Upsilon$ that are able to fit the entire range of data using both dark and visible components, as shown in Figure \ref{curvegrid}. The free $f_\Upsilon$ run gives $r_1=1.94\pm_{0.4}^{1.76}$ (90\% confidence), which has an upper bound outside the data range. The lower bound roughly corresponds to the point where the contribution to the circular velocity from the dark matter halo first exceeds that of the stellar disk.


\subsection{NGC 3198}


This galaxy has a well-studied rotation curve \citep[e.g. ][]{begeman1989}. Error bars show the disk is roughly axisymmetric, and there is good kinematic data from the HI map at all radii \citep[Figure 75 in ][]{deblok2008}. 

One notable feature of the rotation curve for NGC3198 is that the circular velocity contribution of atomic gas is less than zero in the inner part. This represents a void where the net gravitational effect of the gas disk is outwards, and has to be subtracted in quadrature rather than added. This occurs in the same region as the peak of the inner stellar component in the rotation curve, so if the lack of atomic gas could indicate a large amount of molecular gas in the disk (observations of which are not part of any of the original data sources used here) then the contribution of such gas can be modelled by the freedom in $f_{\Upsilon}$. Under this assumption, $f_{\Upsilon}$ no longer functions purely as a factor of stellar mass-to-light ratio.

NGC 3198 is another high surface brightness galaxy, but this itself does not preclude a constraint. The initial run produced differing constraints for each chain, as did a run using the higher temperature setting $T=2$, but we were able to produce a consistent constraint by also excluding those data points where the neutral gas contribution is negative - which also includes the supermaximal inner stellar component. This constraint had a $\gamma_{\rm}$ peak greater than 1, and thus no meaningful value for $r_1$. 

\begin{figure}
  \includegraphics[width=\linewidth]{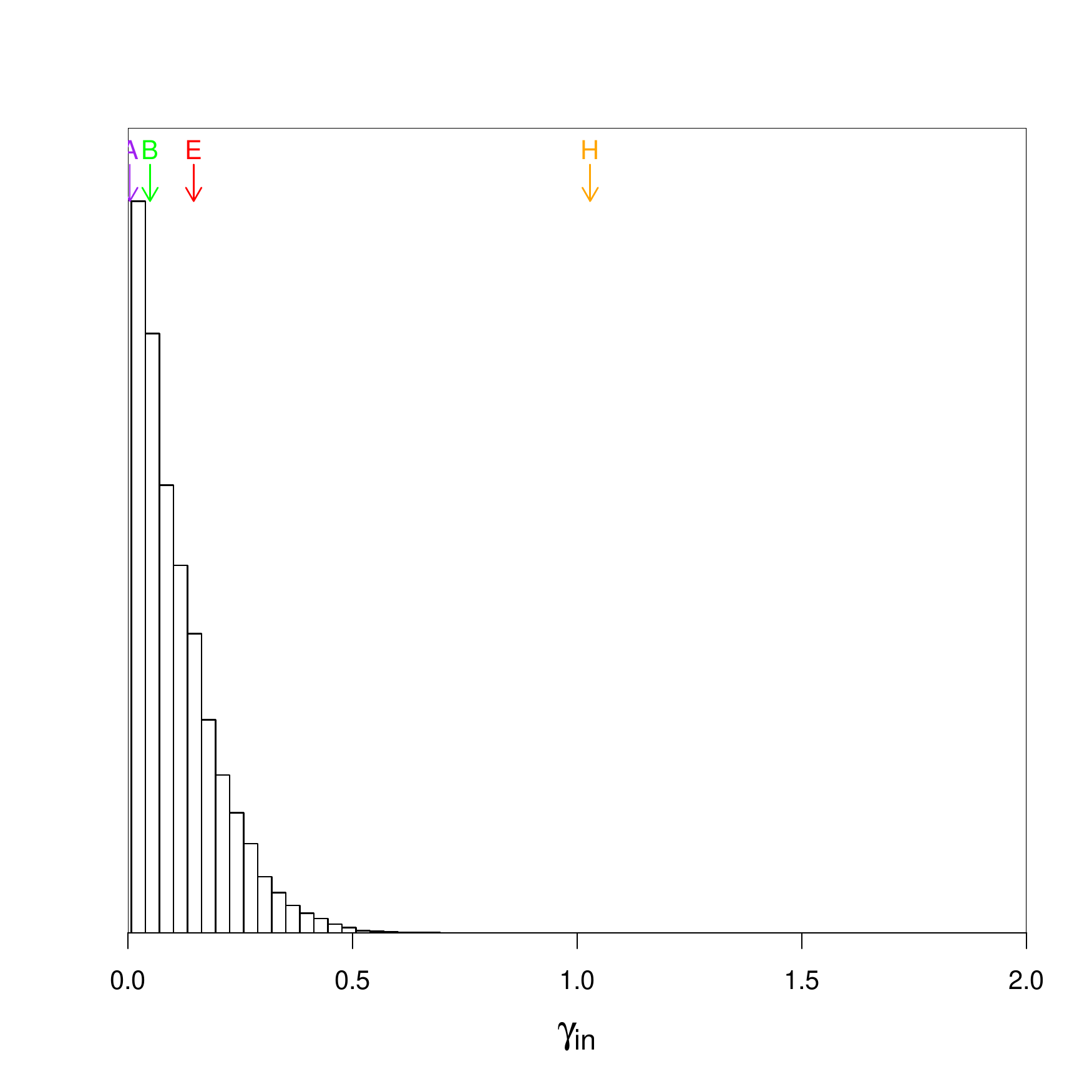}
  \includegraphics[width=\linewidth]{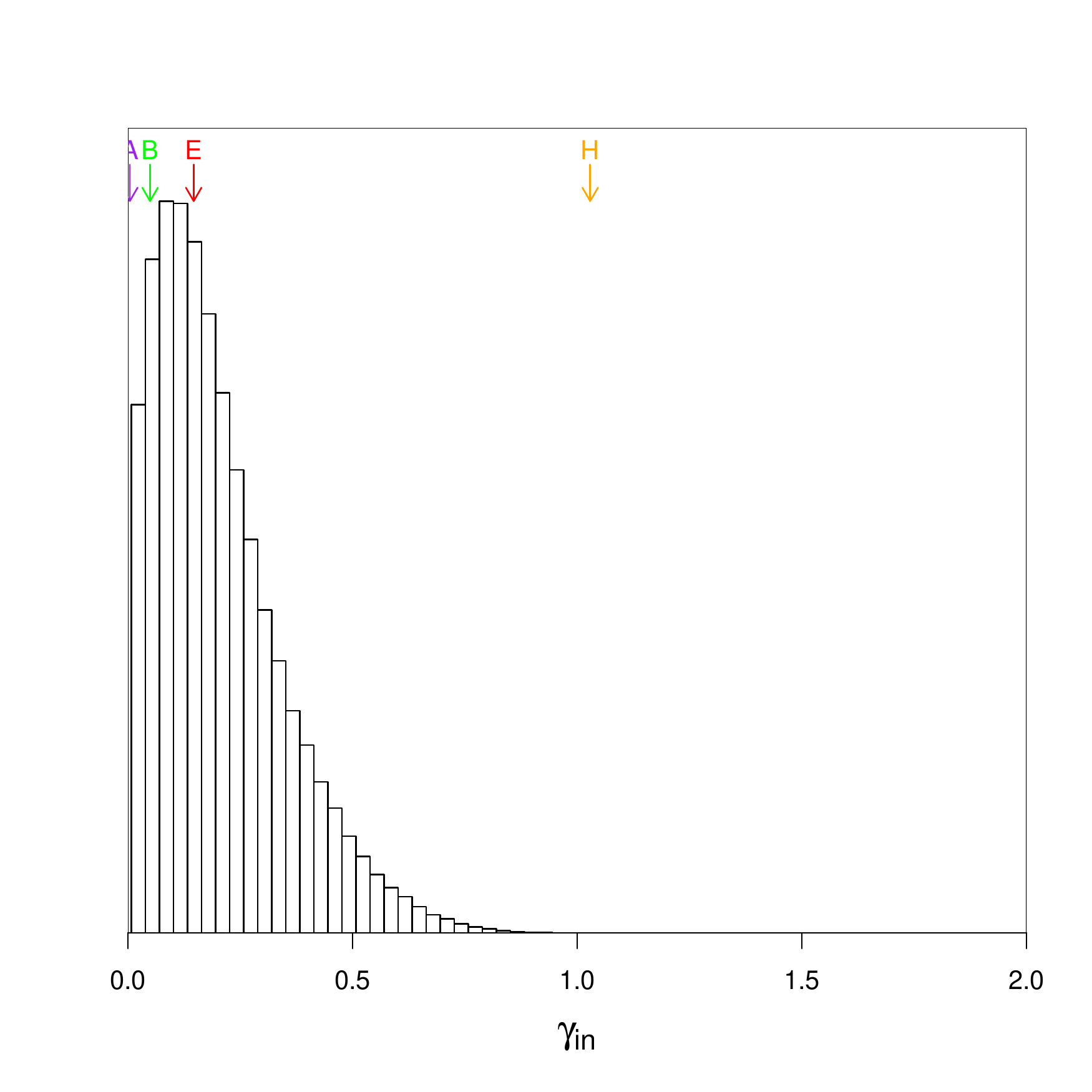}
  \caption{Histogram of the $\gamma_{\rm in}$ value of all the models produced by the NGC 3198 MCMC run using fixed $f_{\Upsilon}$. {\bf Top} Results derived from a Kroupa IMF and {\bf bottom} results derived from a diet Salpeter IMF.}
  \label{NGC3198results}
\end{figure}

Constraints at $T=1$, using the entire data range, were possible using a fixed $f_{\Upsilon}$, with the results shown in Figure \ref{NGC3198results}. As these both produce constraints on $\gamma_{\rm in}$ that agree to within $1\sigma$, we classify this galaxy as having a core, but this assumes that the stellar mass-to-light modelling is robust. The models produced by the free mass-to-light multiplier run are almost entirely ($>88\%$) below $f_{\Upsilon}=0.341$, indicating that the MCMC prioritises keeping the stellar contribution low in order to keep the inner part of it lower than the observed circular velocity of the galaxy. Better stellar mass modelling is required for a more definitive constraint.

\subsection{NGC 3521}



We used a temperature setting of $T=2$ for this galaxy as the lower temperature run produced a constraint value of $\sigma(\hat{x})/\hat{\sigma}(x) = 2.91$ for $\beta$. Due to the nature of the data, and the area of our interest being the central region of the galaxy, having a poorly constrained outer log slope is not in itself grounds for a rejecting a result, but in this case both $\alpha$ and $r_{\rm s}$ also had  $\sigma(\hat{x})/\hat{\sigma}(x) > 1.0$ when $T=1$, so we only present the higher temperature run here. 

 \begin{figure}
  \includegraphics[width=\linewidth]{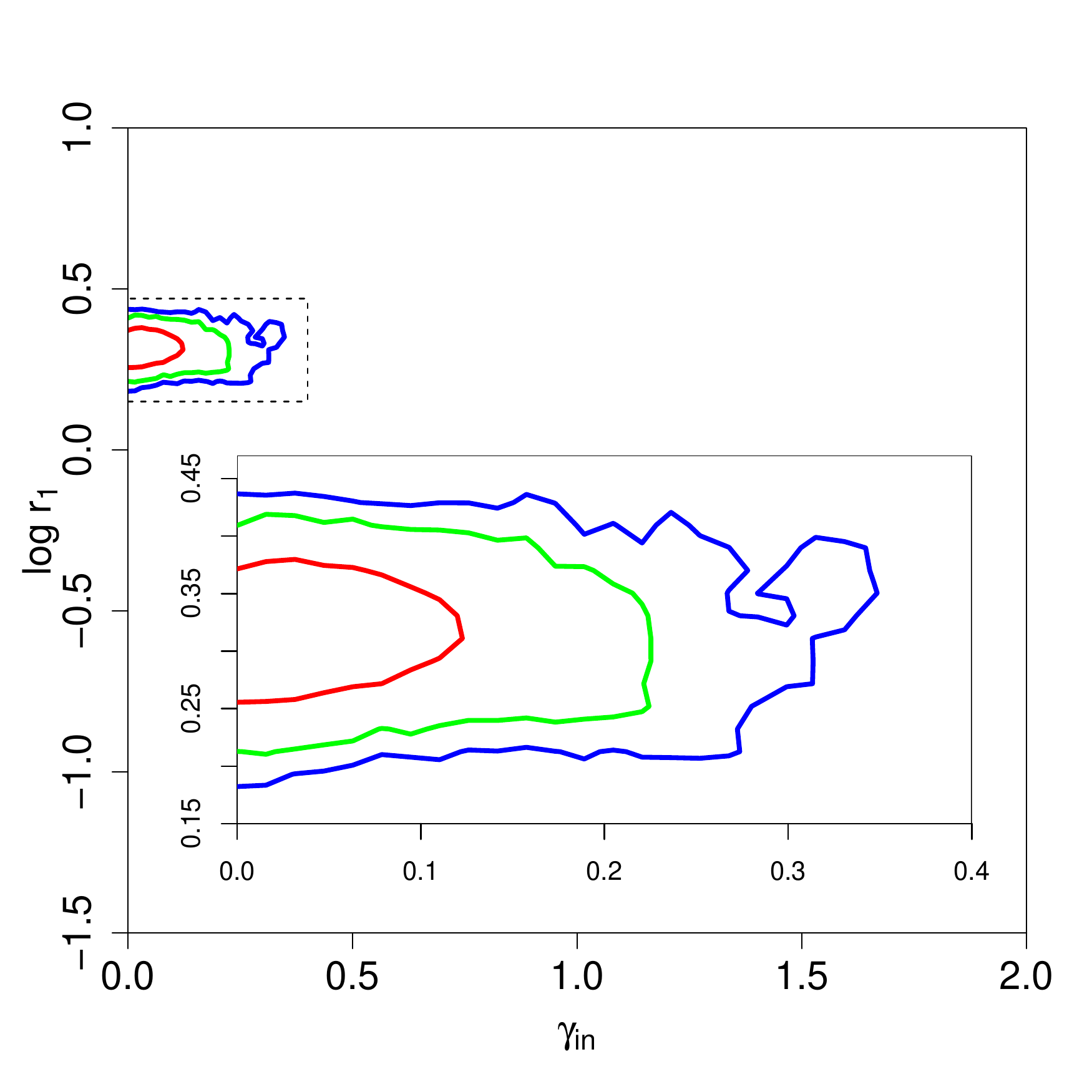}
  \caption{Contour plot of log $r_1$ versus $\gamma_{\rm in}$ for NGC 3521. Inset shows a zoom in on the contoured region.}
  \label{NGC3521gammainr1}
\end{figure}

\begin{figure}
  \includegraphics[width=\linewidth]{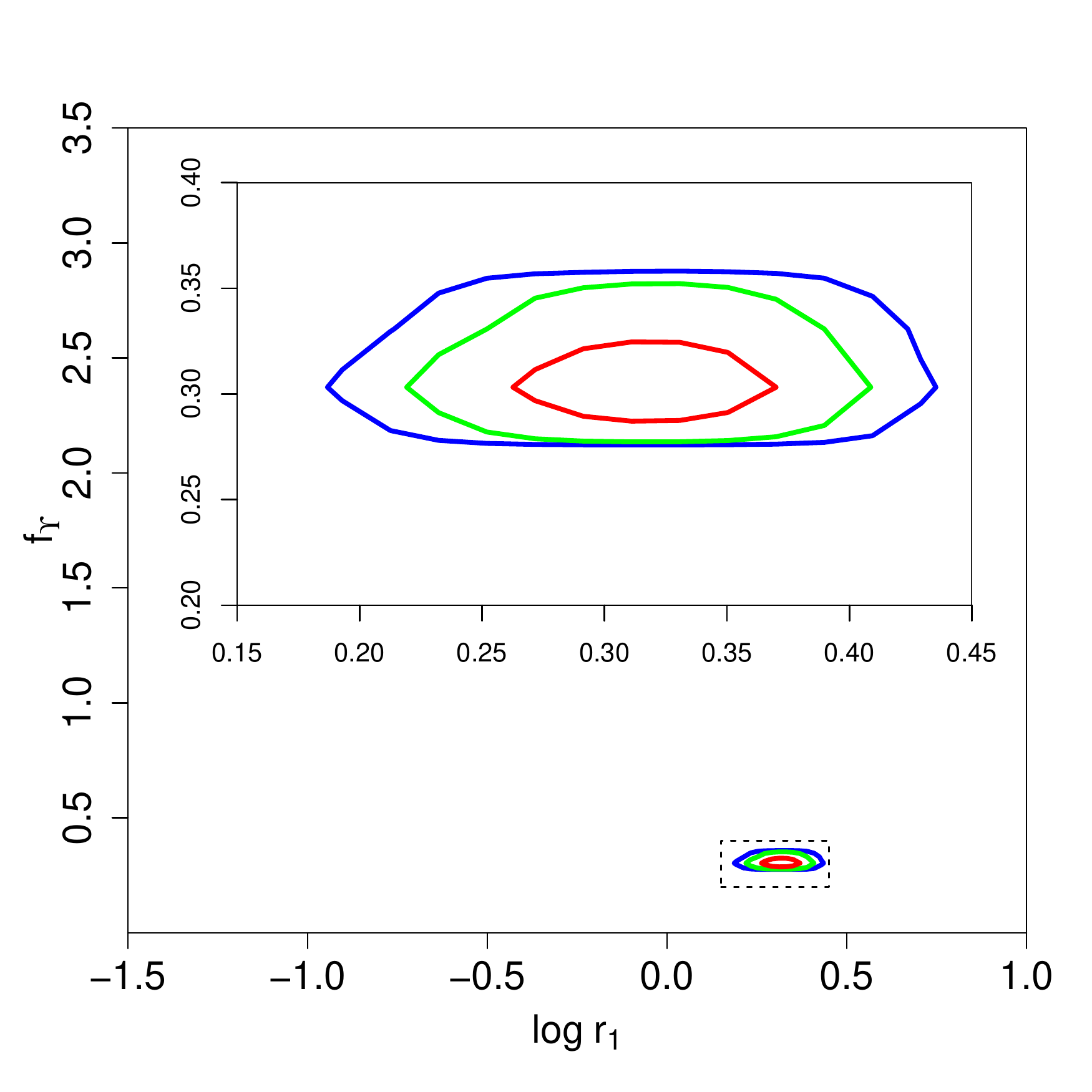}
  \caption{Contour plot of the mass-to-light multiplier $f_\Upsilon$ versus $r_1$ of all the models produced by the NGC 3521 MCMC run.}
  \label{NGC3521r1ml}
\end{figure}

A preference for a core is obtained, as shown in Figure \ref{NGC3521gammainr1}, being flat to within reasonable error at the innermost data point. This remains unchanged regardless which of our initial assumptions about stellar mass-to-light is used, although in the cases of fixed IMF the dark matter contribution is negligible at the smallest radial bin ($r=312{\rm pc}$.) A strong constraint is however produced on $r_1$ as shown in Figure \ref{NGC3521r1ml}. There is no dependency of $r_1$ on $f_\Upsilon$,  which is also strongly constrained, in the free mass-to-light case. However, the exact value that $r_1$ is constrained to changes if we use a fixed $f_\Upsilon$ based on a Kroupa or diet Salpeter IMF. This is discussed further in Section \ref{discussion}.

The rotation curve decomposition for NGC 3521, shown in Figure \ref{curvegrid}, indicates a gap in the neutral atomic gas disk and also a stellar contribution that is larger than the total rotation curve for the input value of $f_{\Upsilon}$, corresponding to a diet Salpeter IMF. Only using the data in the range where the neutral gas contribution to the rotation curve is positive would not permit modelling of the density profile shape as all the contributions and the observed rotation curve in the remaining region is almost flat. The rising part of the dark matter halo contribution to the rotation curve is required in order to differentiate between a central core and a cusp. An MCMC run excluding the data points where the velocity of the gas contribution is negative produced a peak $\gamma_{\rm in} \geq 2$, but as stated above this cannot be considered a meaningful value.

\subsection{NGC 3621}

This galaxy has a stellar disk and a rotation curve of similar maximum velocity and shape as that of NGC 3198, and like that galaxy has a high surface brightness. However, the galaxy exhibits significantly more of a cusp when analysed. 

\begin{figure}
  \includegraphics[width=\linewidth]{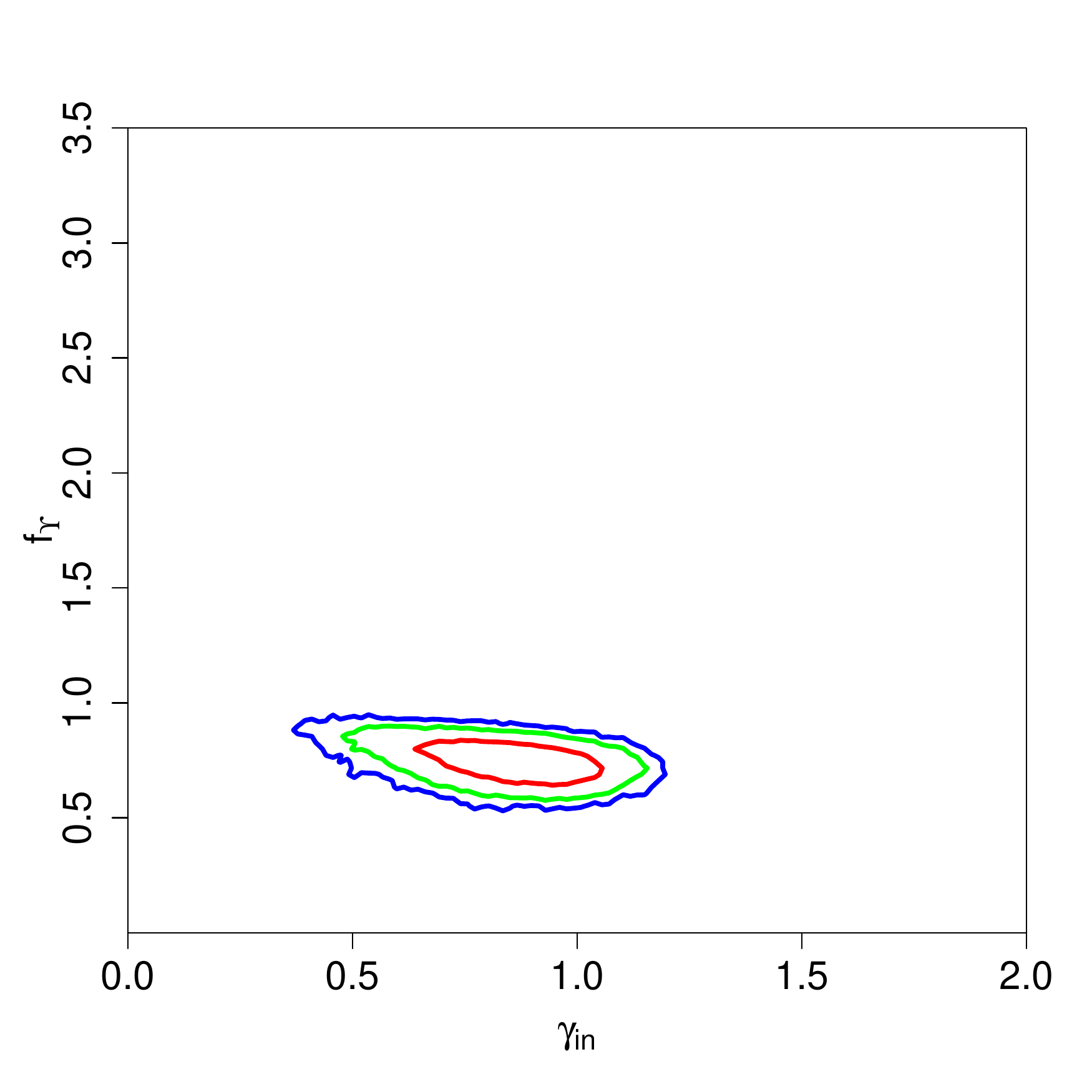}
  \caption{Contour plot of the mass-to-light multiplier $f_\Upsilon$ versus $\gamma_{\rm in}$ of all the models produced by the NGC 3621 MCMC run. }
  \label{NGC3621gammainml}
\end{figure}

In Figure \ref{NGC3621gammainml} we see there is a slight degeneracy between $\gamma_{\rm in}$ and $f_\Upsilon$. Although $\gamma_{\rm in}$ is well constrained when the mass-to-light ratio parameter varies freely (the red contour on the map showing the $1\sigma$ level, with an overall range $\gamma_{\rm in} = 0.91\pm^{0.13}_{0.33}$), or when the ratio is fixed to a particular value, the position of the distribution varies from moderately cored ($\gamma_{\rm in} = 0.33\pm^{0.31}_{0.16}$ when using a diet Salpeter IMF) to almost entirely cusped ($\gamma_{\rm in} = 0.89\pm^{0.16}_{0.2}$ when using a Kroupa IMF.)

The cusped interpretation is favoured when the MCMC can control $f_\Upsilon$, however this distribution is a product of our prior assumptions. The correct value of $f_\Upsilon$ also needs to account for any molecular gas not included in the model, so it is difficult to give a precise value for it. We cannot at this point state which of the $\gamma_{\rm in}$ values is correct from just this analysis.


\subsection{NGC 7793}

NGC 7793 is another high surface brightness galaxy, but at first sight we appear to be able to constrain it reasonably well. We generated a test case for high surface brightness galaxies based on NGC 7793 in HW13. 

\begin{figure}
  \includegraphics[width=\linewidth]{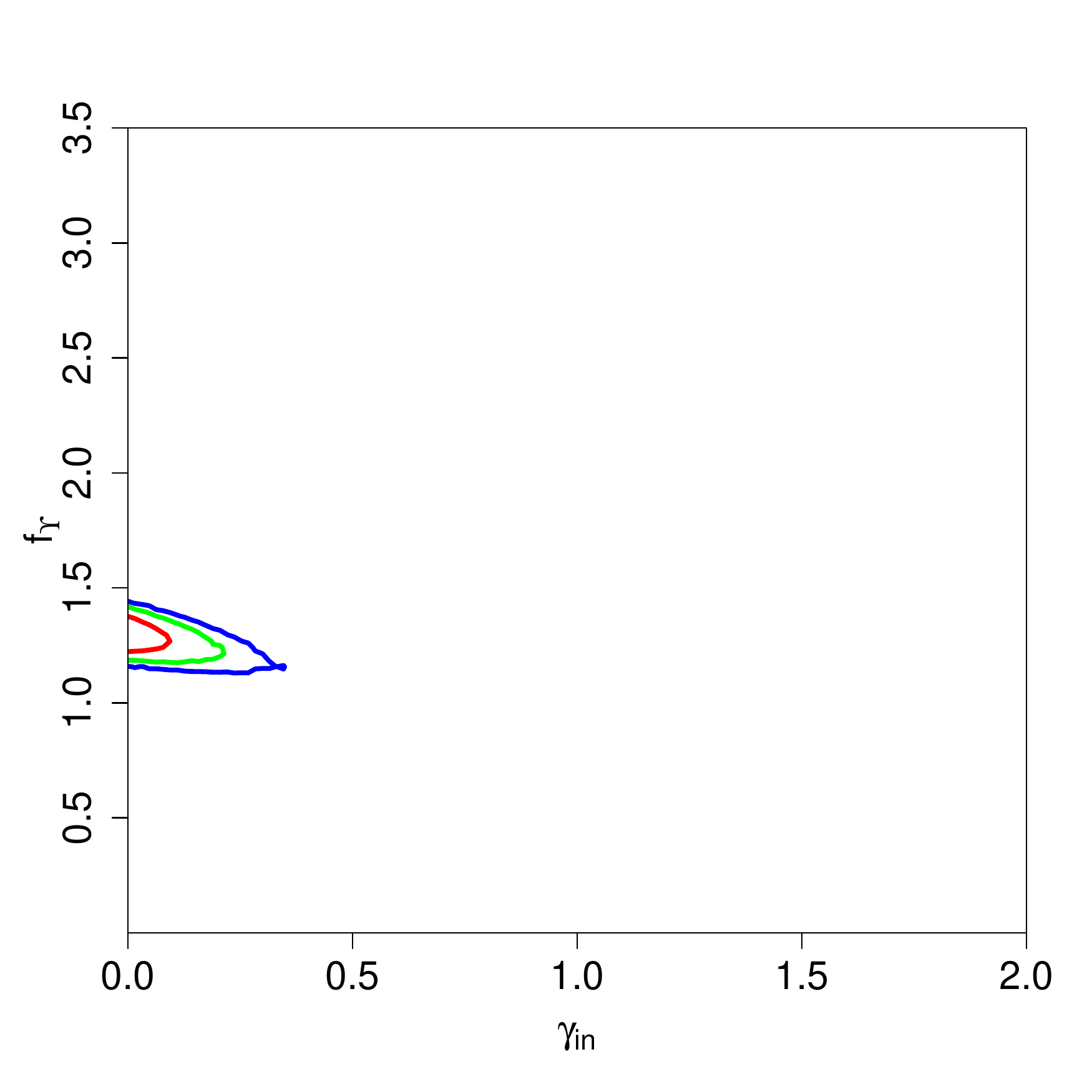}
  \caption{Contour plot of the $\gamma$ value versus mass-to-light multiplier $f_\Upsilon$ of all the models produced by the NGC 7793 MCMC run. }
  \label{NGC7793gammainml}
\end{figure}

\begin{figure}
  \includegraphics[width=\linewidth]{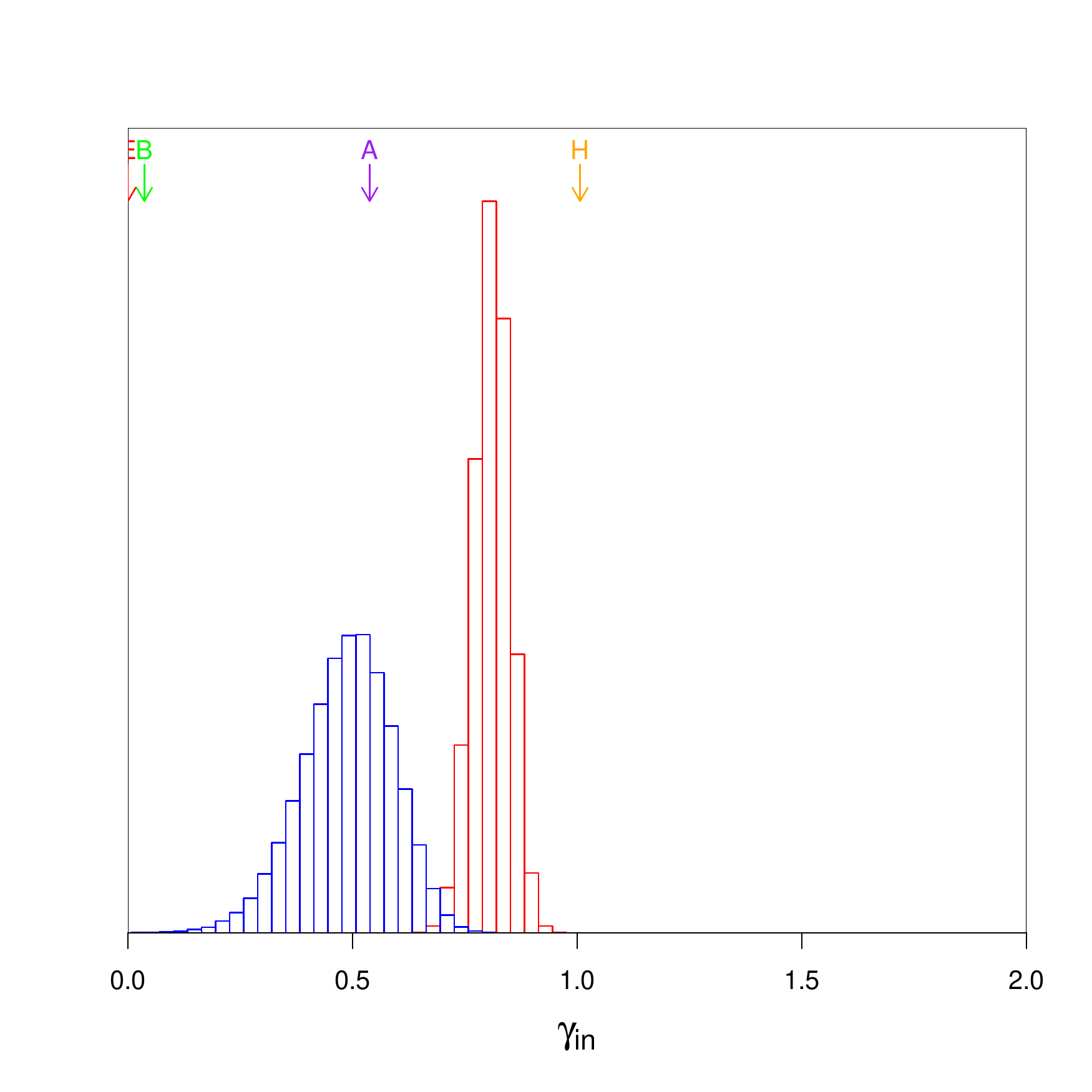}
  \caption{Histograms of $\gamma_{\rm in}$ values for NGC 7793 runs \textbf{blue} using a diet Salpeter IMF derived mass-to-light ratio, and \textbf{red} using a Kroupa IMF \protect\citep{kroupa2001} derived mass-to-light ratio. Histograms are scaled to have equal integrated area. For a Kroupa IMF, $f_\Upsilon=0.72$ and for a diet Salpeter IMF $f_\Upsilon=1$.}
  \label{NGC7793double}
\end{figure}

According to Figure \ref{NGC7793gammainml}, both the inner log slope and the mass-to-light multiplier $f_\Upsilon$ are well constrained and show a clearly cored profile. However, the constraint indicates a mass-to-light ratio which is substantially higher than that implied by a diet Salpeter IMF, possibly due to the involvement of molecular gas. Different values, with equally tight constraints, are found when a fixed mass-to-light ratio is chosen, as shown in Figure \ref{NGC7793double}

This demonstrates a reason why the likelihood distributions in Figure \ref{NGC7793double} must be used with caution. Each distribution, taken in isolation, returns a very strong constraint - but it is only apparent through a broader analysis of the parameter space of the result that this constraint is entirely dependent on the initial assumption about the stellar mass-to-light ratio. This emphasises the value of the MCMC approach.

\begin{figure}
  \includegraphics[width=\linewidth]{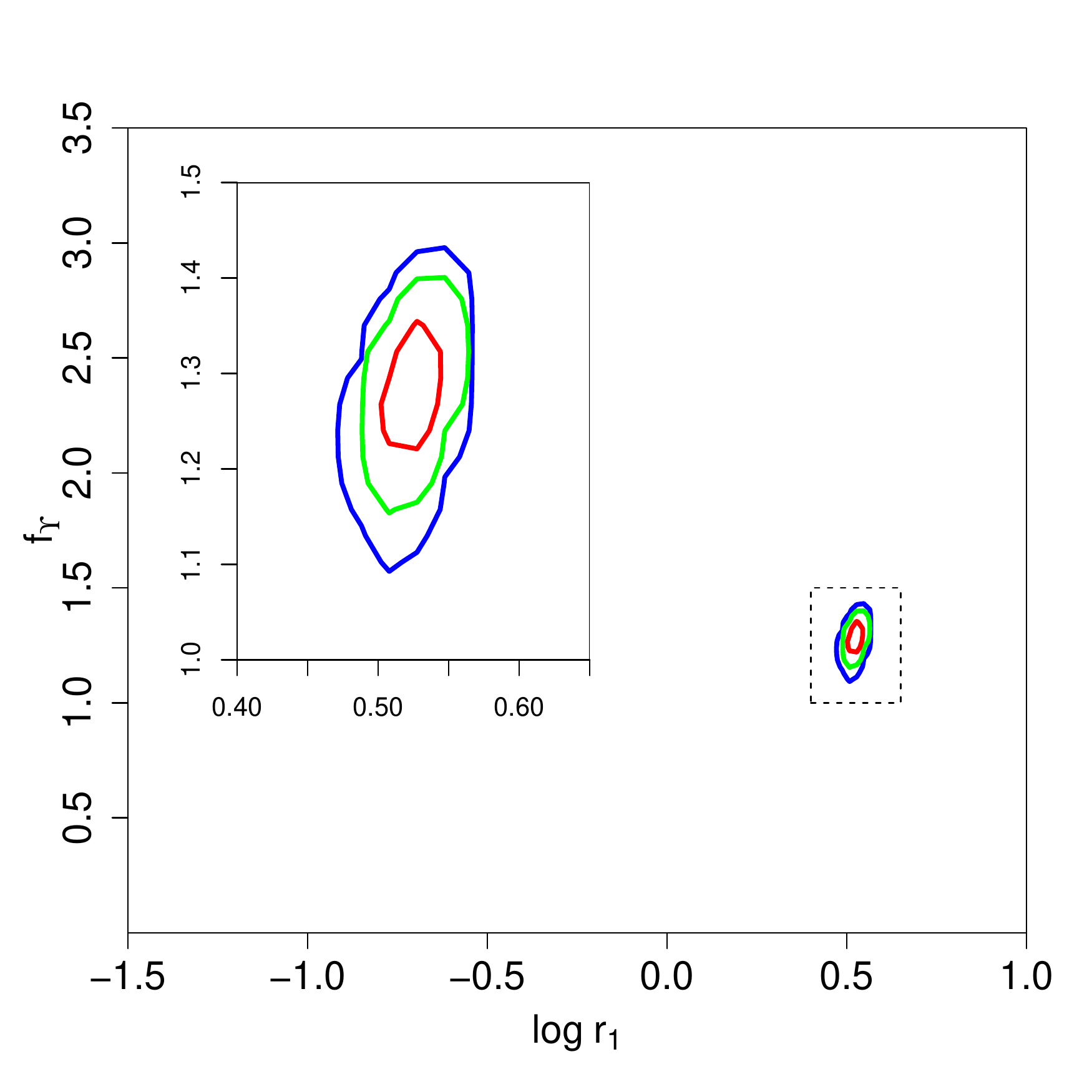}
  \caption{Contour plot of $f_\Upsilon$ versus $r_1$ for NGC 7793. Inset shows a zoom in on the contoured region.}
  \label{NGC7793r1ml}
\end{figure}

\subsection{NGC 925}

This galaxy is described in \cite{deblok2008} as having a weak bar, which is found in \cite{elmegreen1997} to have a length of $5.4{\rm kpc}$. The impact of this bar on the kinematics, and thus the rotation curve, of the galaxy must be taken into account in our analysis. We do this through two separate runs of this galaxy, one with the entire data range from \cite{deblok2008}, and one excluding the data points inside $r=5.6{\rm kpc}$ in order to minimise the impact of bar kinematics on our result.

In a previous analysis \citep{chemin2011}, the calculated rotation curve of this galaxy was found to be well fit by an Einasto profile dark matter halo that featured a sharp change in the rotation curve, caused by the single Einasto shaping parameter ($n$ in \citealt{chemin2011} but also referred to as $\alpha$ in \citealt{navarro2004}) being very low. Such a halo was avoided in our analysis through the availability of more shaping parameters.

\begin{figure}
  \includegraphics[width=\linewidth]{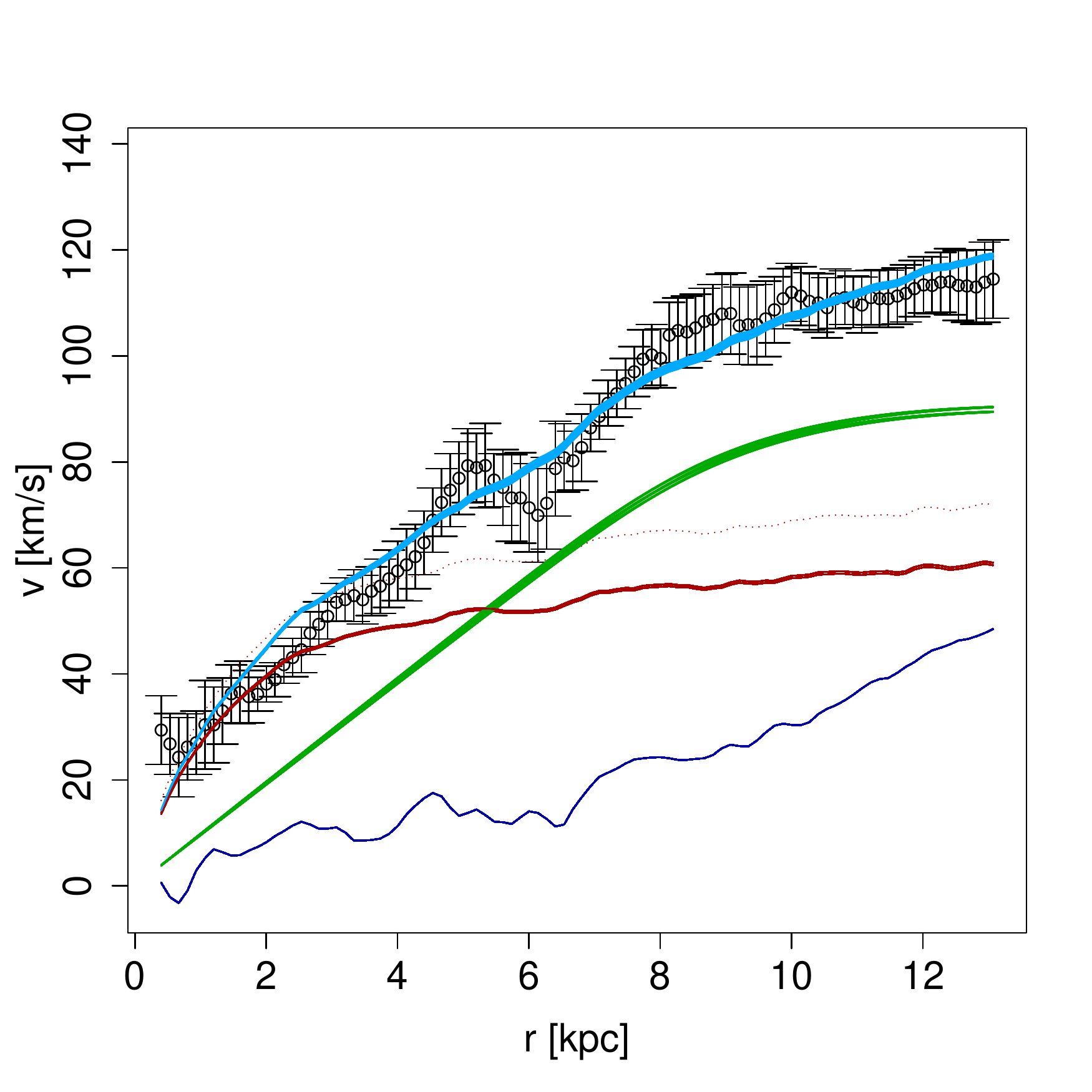}
  \caption{Rotation curve produced by the MCMC analysis of the data for NGC 925 assuming a fixed Kroupa IMF. The black data points are HI rotation curve data, the red solid line is the modelled stellar contribution derived from a Kroupa IMF to the rotation curve, the dark blue line is the modelled gas contribution, the green line is the dark matter halo model at the peak of the distribution, and the light blue line is the expected rotation curve produced by all these components. For comparison the dotted line shows the stellar component assuming a diet Salpeter IMF, which is not used in this calculation.}
  \label{NGC925curve1}
\end{figure}

The fits to this halo strongly point to the existence of a core, with a radius that encompasses most of the data range. Forcing a fixed mass-to-light multiplier does not change the result, it merely reduces the size of the tail of the distributions. Assuming a diet Salpeter IMF results in a supermaximal disk i.e. the calculated stellar contribution being higher than the observed rotation velocity, for a number of points towards the inner part of the rotation curve, so we conclude that the Kroupa IMF derived stellar component represents a more realistic fixed $\Upsilon$.

As shown in Figure \ref{NGC925curve1}, the contribution of the stellar component is high at small $r$, but the surface brightness becomes much lower further out in the disk. This could mean that the cored profile (which has the steepest rotation curve) is simply the one which allows the dark matter contribution to most quickly transition from almost irrelevant, to being the dominant contribution, as radius increases. There is also a feature at around $r=5{\rm kpc}$ that is not modelled well by either the baryons or the proposed dark matter halo, and may be related to the bar.

The above does not prevent our result being robust. The cored portion of the proposed dark matter halo extends from where the stellar contribution stops matching the shape of the observed rotation curve, through to where the dark matter contribution is dominant. So, whilst the innermost data points may not able to constrain a core, if there were not a flat density profile at these points, the overall profile would be surprising as it would have uniform density at intermediate radii, and a rising density again interior to this.

Due to the influence of the bar, and the fact that the rotation curve indicates a larger stellar contribution than that of dark matter at small $r$, we do not consider the log slope here itself to be evidence of a cored density profile. However, we note that analysis of this galaxy gives a value of $r_1 = 6.84 \pm ^{0.41}_{0.53}$ (90\% confidence interval) which places the scale radius beyond the radial extent of most of the bar. 

\begin{figure}
  \includegraphics[width=\linewidth]{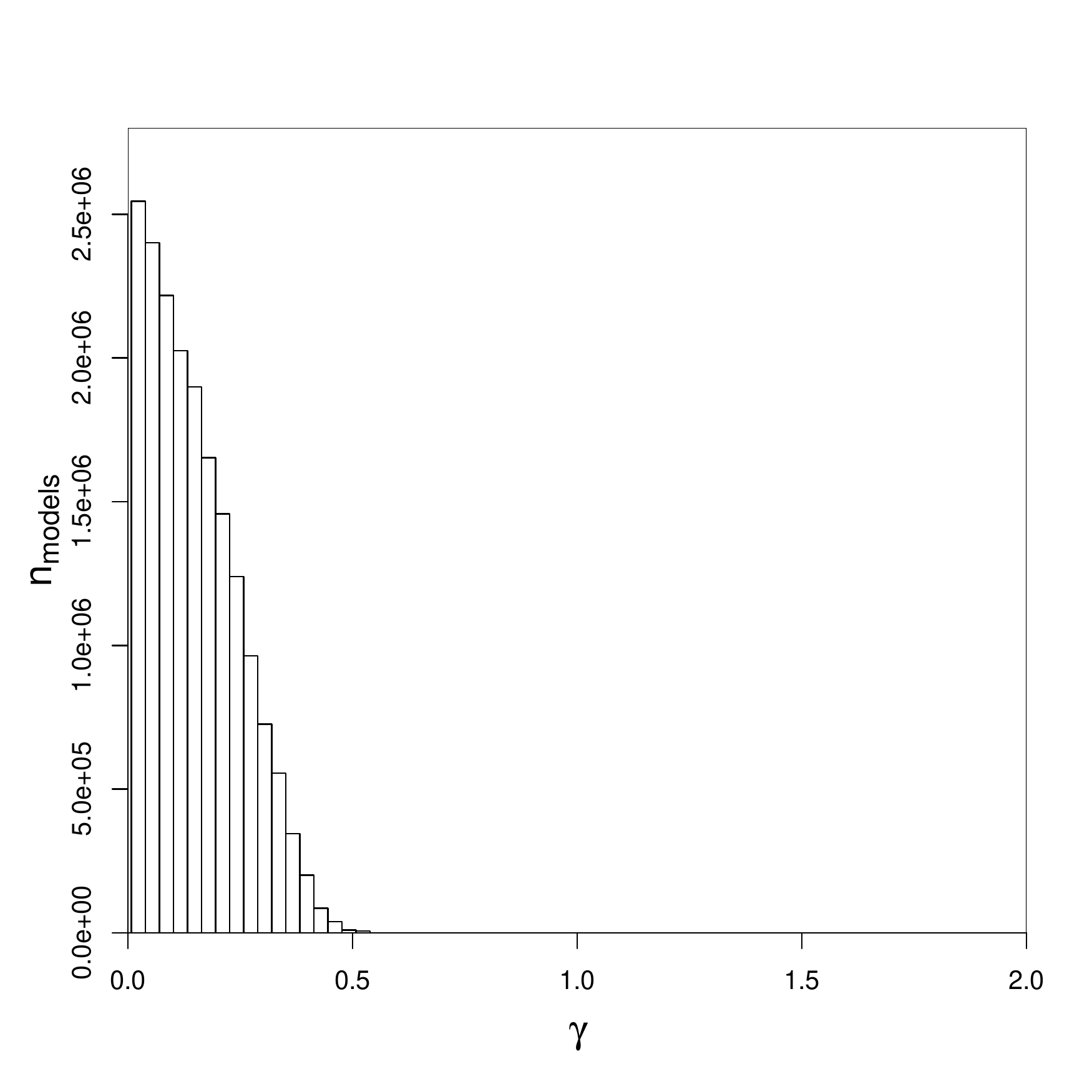}
  \includegraphics[width=\linewidth]{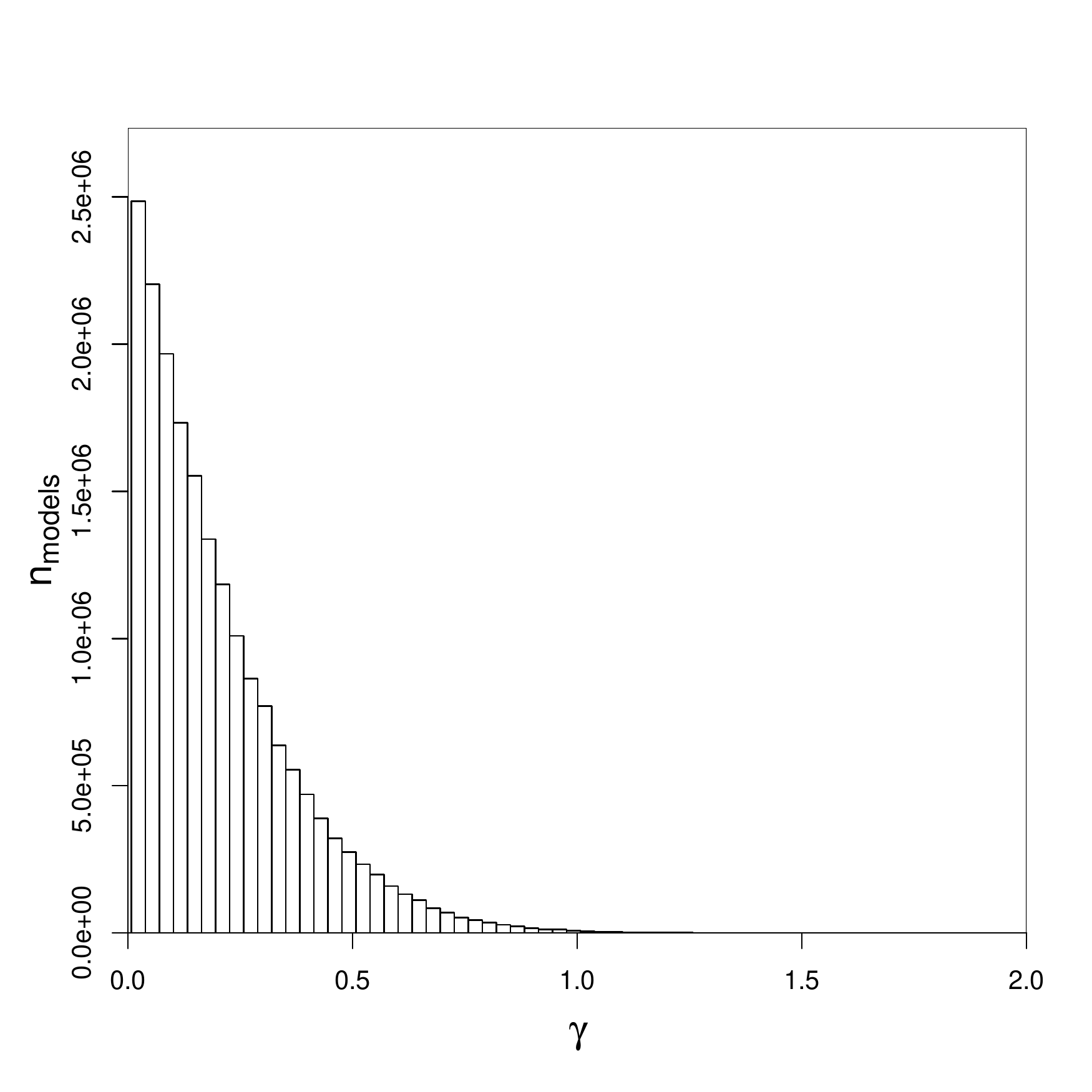}
  \caption{Distributions of $\gamma$ for NGC 925 using all data points ({\bf top}) and  using only those data points outside $r_1 = 6.77$ ({\bf bottom}). Note that we do not use $\gamma_{\rm in}$ for this comparison as the two cases have different inner radial bins. See text for a discussion.}
  \label{NGC925compare}
\end{figure}

Our second run, excluding the inner part of the rotation curve, showed $r_1 =  6.77 \pm ^{0.65}_{0.75}$. In this case $\gamma$ is less well constrained, as illustrated in Figure \ref{NGC925compare}. This shows that for this galaxy the $r_1$ result is not compromised by any relation between $\gamma$ and $r_1$. We conclude that $r_1$ is well constrained for this galaxy, but $\gamma_{\rm in}$ is much less well constrained. However, it should be noted that higher values of $\gamma_{\rm in}$, associated with a cusped profile, are much more difficult to reconcile with the value we have for $r_1$ as it would imply a very slow change in log slope and thus provide a poor fit at larger $r$.

Our modelling can also explain the anomalous result obtained for this galaxy by ~\cite{chemin2011}. The rotation curve requires a flat log slope over a large radial range in order to produce a good fit, and the only way to do this with an Einasto halo is to lower the shape parameter $n$ radically, which also leads to an unphysical, sharp drop in density near $r = 10{\rm kpc}$. A profile with more parameters (in this case, $\gamma$ controlling the inner log slope and $\alpha$ controlling the rate of transition from the inner to the outer asymptotic log slopes) avoids this problem. 

\subsection{Rejected Galaxies}

We attempted to apply this technique to NGC 2841, NGC 2903, NGC 3031, NGC 4736, NGC 5055 and NGC 7331, but we found that there was either inadequate kinematic data for the MCMC algorithm to find a genuine fit to the data, or there was sufficient underlying asymmetry in the disk to prevent the MCMC method finding a useful constraint.

In the case of NGC 5055, the morphology of the galaxy and the shape of the rotation curve both appear promising as a target for this technique. However, when we applied an MCMC analysis, we were unable to constrain any parameters. In Figure \ref{curvegrid} we showed the peak of the distribution. The halo corresponding to the most populated bin in the parameter space fits the data well, and relying purely on the value of $\chi_{\rm red}^2 < 0.5 $ without the context of the parameter space, the conclusion would be that the halo has been correctly modelled. 

\begin{figure}
  \includegraphics[width=\linewidth]{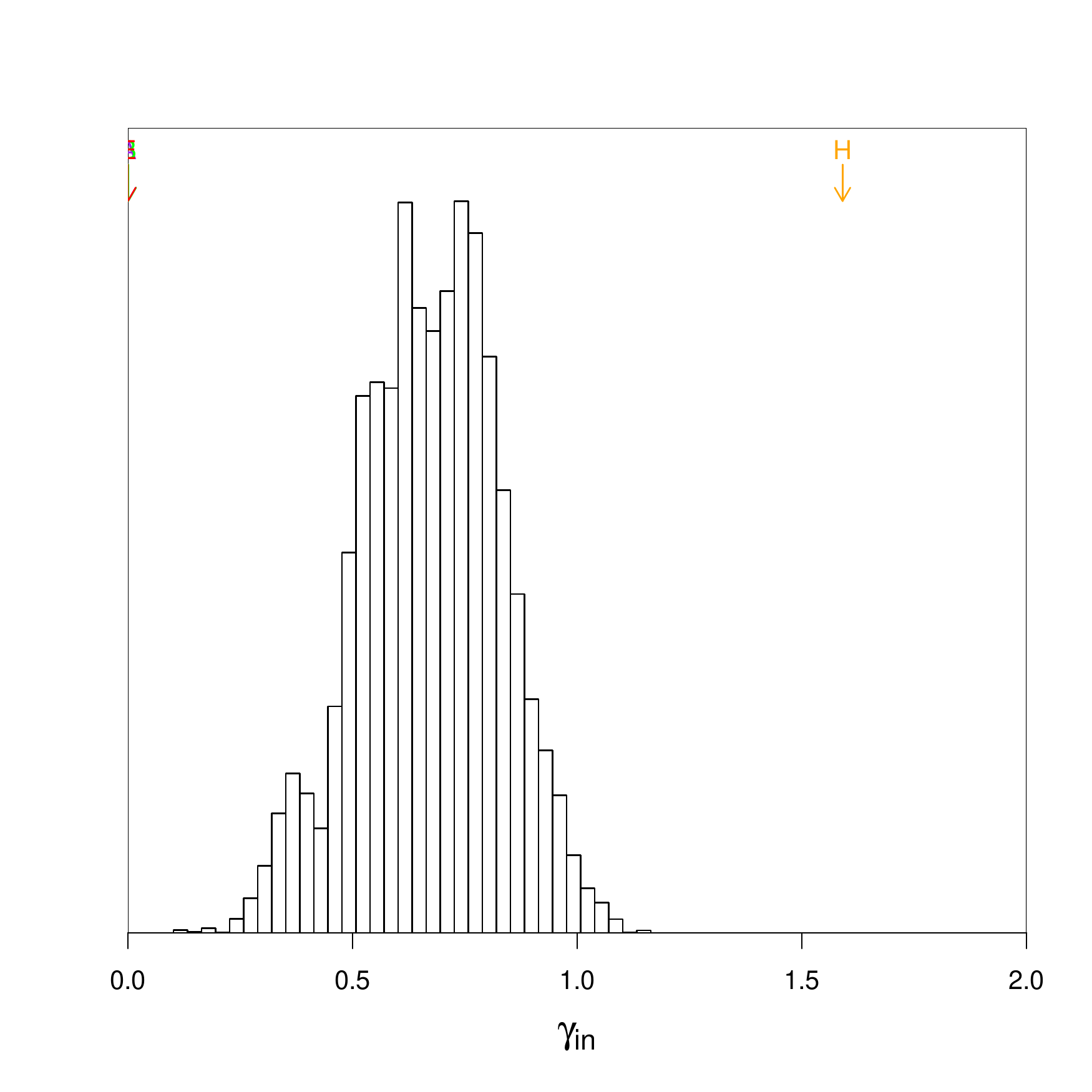}
  \caption{Histogram of $\gamma_{\rm in}$ values from an MCMC run on NGC5055 with a temperature setting $T=1$. There are $\sim4 \times 10^7$ models included from 8 independent chains.}
  \label{NGC5055}
\end{figure}

The extra information provided by the MCMC process allows us to show that this is not the case. In Figure \ref{NGC5055} we see the distribution of $\gamma_{\rm in}$ values is not smooth over a length scale comparable to the initial step size of the MCMC chain, suggesting the 8 parallel chains have not converged to the same distribution, despite them visiting over $\sim4 \times 10^7$ models between them. At temperature settings $T$=1, 2, and 3 we were unable to produce a set of chains with a convergence statistic $\sigma(\hat{x})/\hat{\sigma}(x) < 1$ for all parameters. Without a repeatable probability distribution, we cannot draw any conclusions.  

\section{Discussion}
\label{discussion}
\label{discussionsection}

\begin{figure}
  \includegraphics[width=\linewidth]{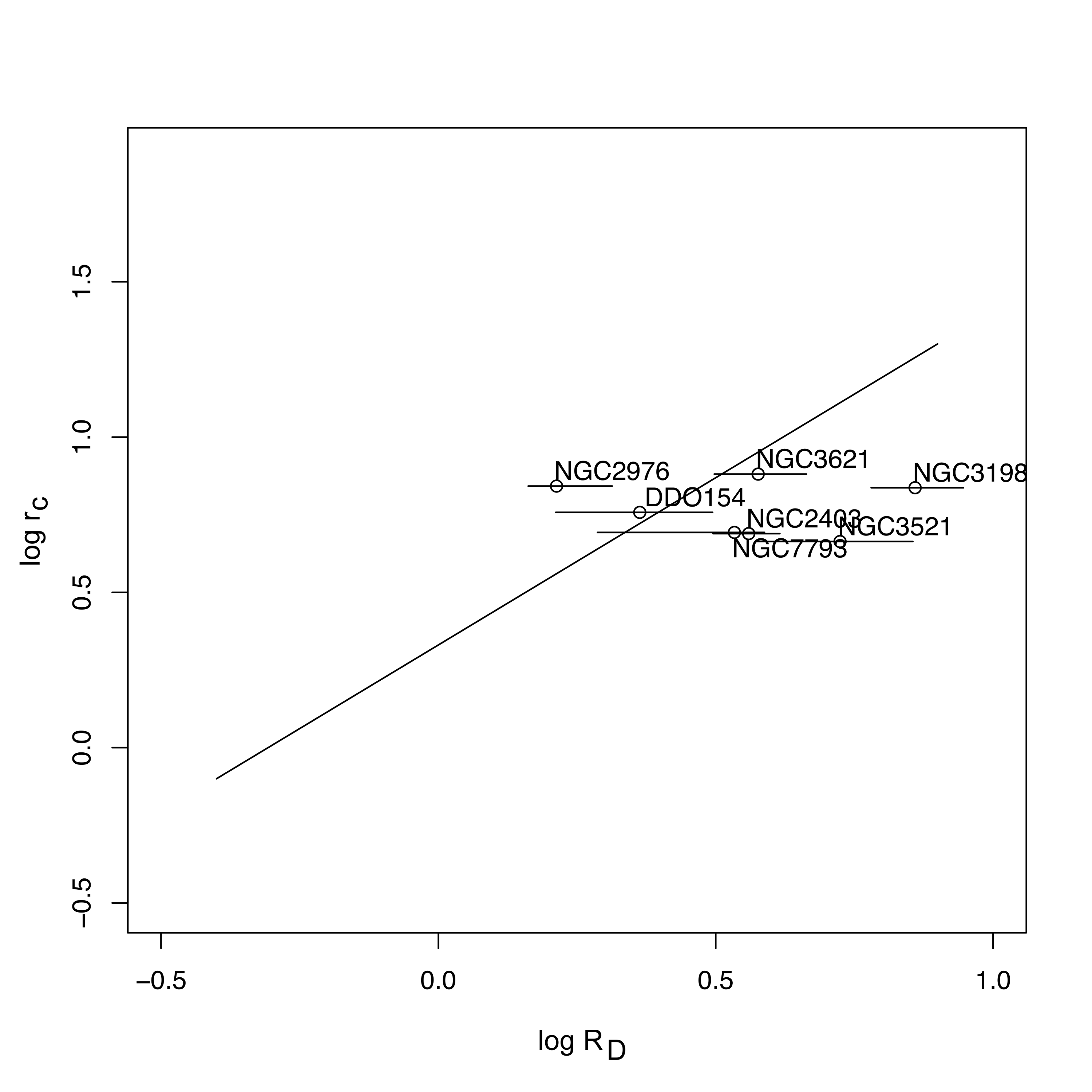}
  \caption{The relation between the dark matter core radii, as defined in equation \ref{piso} and stellar disk radii for our set of galaxies assuming a cored psuedo-isothermal halo profile. NGC 925 is off the right hand edge of this plot. The overlaid solid line is the relation identified by \protect\cite{donato2004}.}
  \label{corrplot}
\end{figure}

\begin{figure}
  \includegraphics[width=\linewidth]{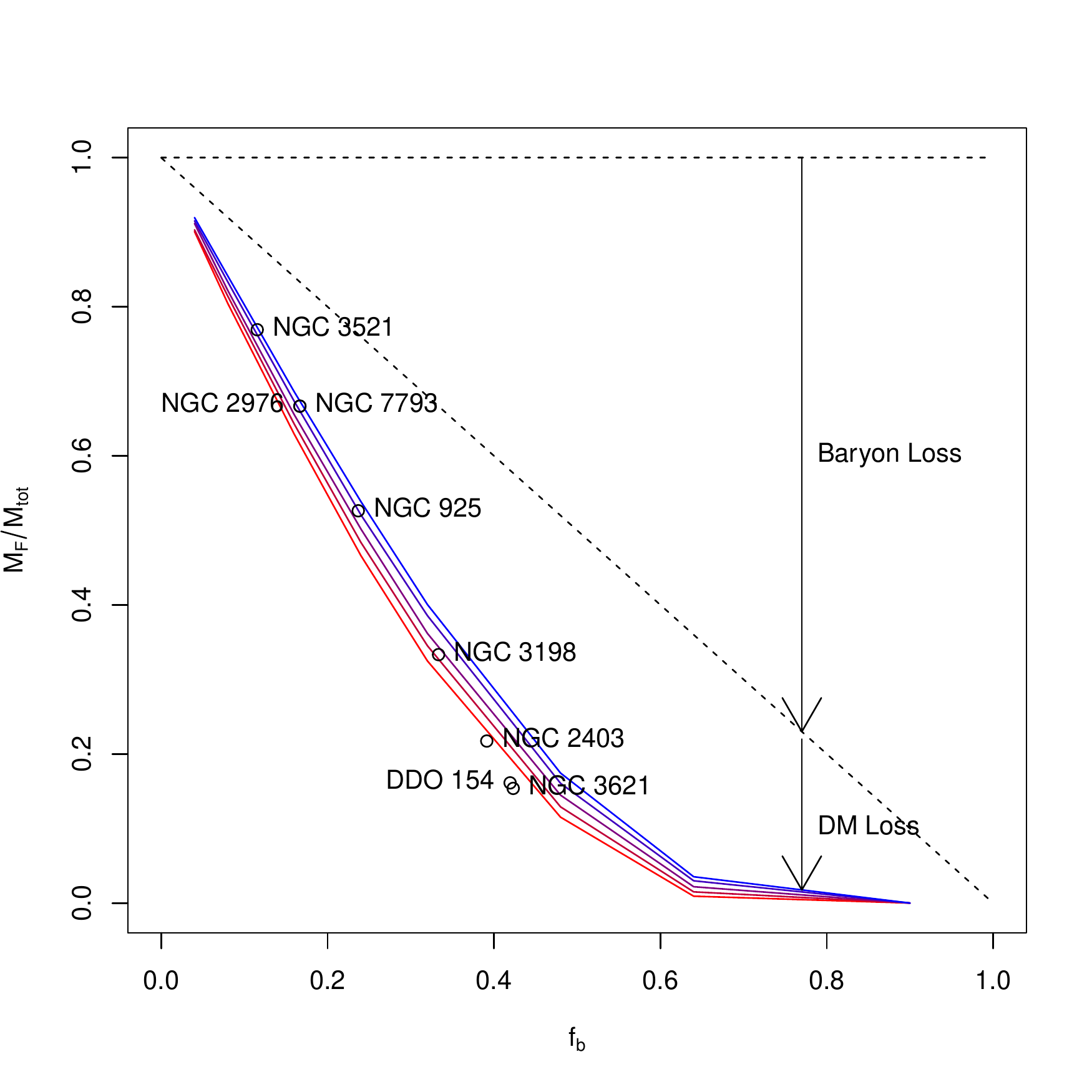}
  \caption{Comparison of data from Table \ref{fbtable} with simulated predictions. $M_{\rm tot}=M_{\rm I} + M_{\rm b}$ is the total mass, baryons and dark matter, interior to $r_1$. The dotted line shows the remaining mass after baryon removal, and the solid lines show the overall mass loss after 1Gyr for various halo profiles. As in Figure \ref{efficiency}, red to blue denotes increasing log slope (0, 0.25, 0.5, 7/9, 1). Simulations were run for baryon fractions $f_b=0.02, 0.04, 0.08, 0.16, 0.24, 0.48, 0.64, 0.90$. The positions of the galaxies are determined by the analytical model, which always assumes equal baryon and dark matter loss and thus they must all lie on in a single line. The simulation results are not by construction forced to agree with the analytical results.}
  \label{masscomp}
\end{figure}

Our MCMC method has produced good constraints on $r_1$ in those galaxies where such a constraint is possible. We now place these estimates in the context of other work, and then use the model we outlined in Section \ref{analysissection} in order to explore what we can deduce about their formation history.

\subsection{Scale length correlation}

In \cite{donato2004} it is suggested that, for a set of 25 galaxies, the scale length of the dark matter halo is proportional to the scale length of the stellar disk. The cored density profile used to argue for proportionality with the disk scale is a pseudo-isothermal halo, given by

\begin{equation}
\rho (r) = \rho_0 {r_{\rm c}^2 \over r^2 + r_{\rm c}^2 }
\label{piso}
\end{equation}

and the stellar component is modelled by an exponential disk with a scale length $R_{\rm D}$. The profile (\ref{piso}) does not show the same degeneracy between $\rho_0$ and $r_{\rm c}$ as the $\alpha-\beta-\gamma$ profile, because it becomes independent of $r_{\rm c}$ at small radii. However it is effectively a single parameter profile. In our analysis we have found that $v_{\rm max}$ is strongly constrained, and this constraint translates into a constraint on $\rho_0 r_{\rm c}^2$ at large radii. Therefore, for any given value of $\rho_0$ (which controls the core behaviour on its own) there is little freedom in $r_{\rm c}$. This in itself does not imply that $r_{\rm c}$ is necessarily meaningless, so we have investigated whether or not there is an equivalent correlation to that found in \cite{donato2004} in the galaxies we are examining. 

\begin{table*}
\begin{tabular}{lccccccccccl}
& \multicolumn{3}{|c|}{Free $f_\Upsilon$} & \multicolumn{3}{|c|}{Kroupa IMF} & \multicolumn{3}{|c|}{diet salpeter IMF} \\
Galaxy & $r_1$ (kpc) & $\gamma_{\rm in}$ & $\chi^2_{\rm red}$ & $r_1$ (kpc) & $\gamma_{\rm in}$ & $\chi^2_{\rm red}$ & $r_1$ (kpc) & $\gamma_{\rm in}$ & $\chi^2_{\rm red}$ \\ \hline

DDO154 & $1.28\pm_{0.28}^{0.41}$ & $0.42\pm_{0.15}^{0.24}$ & 0.2 & $1.18\pm_{0.24}^{0.31}$ & $0.43\pm_{0.14}^{0.23}$ & 0.2 & $1.17\pm_{0.33}^{0.46}$ & $0.5\pm_{0.21}^{0.25}$ & 0.13  \\
NGC2976 & $1.94\pm_{0.4}^{1.76}$ & $0.47\pm_{0.38}^{0.16}$ & 0.26 & $3.24\pm_{0.89}^{1.14}$ & $0.09\pm_{0.07}^{0.26}$ & 0.32 & $<4.18$ & $<0.97$ & 0.58  \\
NGC7793 & $3.3\pm_{0.14}^{0.14}$ & $<0.13$ & 0.52 & $2.93\pm_{0.17}^{0.13}$ & $0.81\pm_{0.07}^{0.06}$ & 0.67 & $3.15\pm_{0.28}^{0.15}$ & $0.51\pm_{0.18}^{0.14}$ & 0.36  \\
NGC2403 & $<2.81$ & $0.72\pm_{0.49}^{0.29}$ & 0.15 & $0.35\pm_{0.25}^{0.15}$ & $0.8\pm_{0.17}^{0.2}$ & 0.29 & $0.75\pm_{0.5}^{0.44}$ & $0.7\pm_{0.26}^{0.34}$ & 0.15  \\
NGC925 & $6.76\pm_{0.51}^{0.46}$ & $<0.29$ & 0.25 & $7.43\pm_{0.4}^{0.45}$ & $0\pm_{0}^{0.06}$ & 0.42 & $8.4\pm_{0.75}^{7.99}$ & $<0.09$ & 0.55  \\
NGC3621 & $<1.44$ & $0.87\pm_{0.26}^{0.18}$ & 0.28 & $<0.88$ & $0.9\pm_{0.2}^{0.17}$ & 0.27 & $4.61\pm_{1.6}^{1.04}$ & $0.39\pm_{0.21}^{0.36}$ & 0.18  \\
NGC3198 & $3.33\pm_{0.97}^{0.14}$ & $0.09\pm_{0.07}^{0.28}$ & 0.38 & $4.19\pm_{0.7}^{0.4}$ & $0.05\pm_{0.03}^{0.3}$ & 0.67 & $6.36\pm_{1.71}^{1.6}$ & $0.14\pm_{0.09}^{0.41}$ & 0.6  \\
NGC3521 & $2.02\pm_{0.26}^{0.36}$ & $0.03\pm_{0.02}^{0.17}$ & 0.68 & $3.88\pm_{1.2}^{1.36}$ & $0.08\pm_{0.04}^{0.38}$ & 1.25 & $30.58\pm_{12.27}^{20.74}$ & $<0.03$ & 2.09  \\

\hline
\end{tabular}
\caption{Table of all calculated $\gamma_{\rm in}$ and $r_1$ values, with 90\% confidence intervals. Reduced $\chi^2$ values shown are estimates of the peak likelihood, taking the 90th percentile of a randomly selected sub-sample of $\sim10^3$ models from the MCMC chains. The $r_1$ value for NGC 3521, assuming a diet Salpeter IMF, is outside of the data range, as are the confidence intervals for all $r_1$ values for NGC 2976. Values for NGC 2403 and NGC 3521 use a temperature $T=2$, sampling $P^{1/2}$ rather than $P$. See text for detailed discussion.}
\label{mcmctab}
\end{table*}
 
In order to exclude the possibility of the reported correlation being an artefact, we generated sets of 100 rotation curves with random, uncorrelated scale radii for the stellar disk and dark halo of $R_{\rm d}=3\pm1$kpc and $r_{\rm s}=5\pm2$kpc respectively. We added Gaussian noise to the data and included $1\sigma$ error bars of the corresponding size. The dark halo models were either Burkert, pseudo-isothermal, or NFW. We then found the best fitting pseudo-isothermal haloes, given a free mass-to-light ratio. In none of the cases did we find a correlation. Give that \cite{donato2004} found that maximal disks were favoured in their models, we then forced a maximal disk to be used before fitting the pseudo-isothermal halo. Again, we did not find a correlation.

We investigated whether there is a comparable correlation from our own data set (shown in Figure \ref{corrplot}), by assuming a maximal disk and fitting a single pseudo-isothermal profile to each galaxy. We do not see a relation as presented in \cite{donato2004}, although we note that our sample is smaller and has a narrower range of properties. It should also be noted that in order to produce maximal disk fits, higher $f_\Upsilon$ values than we allowed in our MCMC modelling had to be assumed in many cases. 

\subsection{Feedback modelling}

The question of whether or not there is a relation between any parameters of the baryonic component of a galaxy and parameters of its dark matter halo is relevant to understanding galaxy formation. It is presumed, based on cosmological models by \cite{navarro1996} and others, that the centres of dark matter haloes begin strongly cusped, and then become cored by some interaction with baryonic matter \citep[e.g.][]{governato2010}.

The $r_1$ parameter we calculate can be used to gain some insight into this. Any process which reshaped the halo must be able to disrupt the potential of the galaxy to at least this radius, under the assumptions of an NFW starting point and a spherically symmetric halo. In Table \ref{mcmctab} we show that for galaxies DDO 154, NGC 2403, and NGC 2976 the values of $r_1$ are for each galaxy within the 90\% confidence intervals of each other for all prior assumptions about mass-to-light ratio used here.  In the case of NGC 7793, the intervals do not entirely overlap but, as Figure \ref{NGC7793r1ml} shows, the constraint on $r_1$ is very strong and not dependent on $f_\Upsilon$ so this is not an issue of baryonic modelling. 

We now focus on DDO 154, the smallest galaxy in our set, because work such as \cite{governato2010} focuses on feedback in dwarf galaxies. Considering supernova feedback, if each 100$M_\odot$ of star formation leads to a single supernova which feeds back $10^{44}$J of energy ($5.6 \times 10^{-6} M_{\rm intial}$ equivalent solar masses), and assuming the current stellar mass is $\approx88\%$ of the initial mass (using a Kroupa IMF gives an initial composition where 12\% of the total stellar mass is O stars \citep{kroupa2001}, and have all undergone supernova), the feedback energy is  $210 M_\odot c^2$. This is larger than the energy change required in the dark matter halo to modify its density profile; the difference in total kinetic energy between DDO 154 and the closest NFW halo (based on least squares fitting) is $1.2 \pm^{1.6}_{0.9} M_\odot c^2$. Energy production is in this galaxy therefore does not constrain this process, so we must focus on how this energy can be transferred to the dark matter halo.

Feedback from a central star forming region in DDO 154 was modelled in \cite{gelato1999} as an attempt to explain the discrepancy between observations of the galaxy and the NFW halo model. They were only able to reproduce the observed rotation curve by assuming a disk more massive than that which is observed in HI emissions, and furthermore \cite{read2005} shows that their method of contracting the dark matter halo cusp could bias its final state after outflow towards being more flat, due to the assumed isotropic velocity structure of the halo. Applying the method described in Section \ref{analysissection} to DDO 154, we find a gas fraction available for feedback $f_{\rm g}=0.42$, compared to the estimate of the currently observed baryon fraction of 0.1 by \cite{carignan1998}, which supports the conclusion that this galaxy requires additional baryonic mass to account for its dark matter halo profile (although it still leaves open the question of where this mass is now). The complete set of values for $f_{\rm g}$, for all the galaxies studied here, are are shown in Table \ref{fbtable}.

Our calculated values for $f_{\rm g}$ are minima, as the contraction of the baryonic component prior to outflow can only increase the dark matter content interior to $r_1$, and thus require a greater amount of gas outflow to remove. If we assume that the process of contracting dark matter through baryon motion has efficiency of order unity for infall as well as for outflow (that is, each unit mass of baryons moved past $r_1$ in either radial direction brings with it a unit mass of dark matter) then significant infall and contraction would move the value of $f_{\rm g}$ towards 0.5. 

In Figure \ref{masscomp} we show the values from Table \ref{fbtable} in the context of the output from the simulations in Section \ref{analysissection}. The construction of the analytical model forces all the galaxies on to one straight line of the plot, so this should not be taken as a physical confirmation of the simulations. The simulations are shown to be consistent one of the main assumptions of the analytical model; that the amount of dark matter removed during an outflow is comparable to the amount of gas removed. 

\begin{table}
\begin{tabular}{lccccl}
Galaxy & $M_{\rm F,dark}/(M_{\rm I,dark} + M_{\rm I,gas})$ & $f_g$ & $\gamma_{\rm in}$ \\ \hline
DDO 154 & 0.16 & 0.42 & $0.42\pm^{0.24}_{0.15}$ \\
NGC 2976 & 0.67 & 0.17 & $0.47\pm_{0.38}^{0.16}$ \\
NGC 7793 & 0.67 & 0.17 & $<0.13$ \\
NGC 2403 & 0.22 & 0.39 & $0.72\pm^{0.29}_{0.49}$ \\
NGC 925 & 0.53 & 0.24 & $<0.29$ \\
NGC 3621 & 0.15 & 0.43 & $0.87\pm_{0.26}^{0.18}$ \\
NGC 3198 & 0.33 & 0.34 & $0.09\pm_{0.07}^{0.28}$ \\
NGC 3521 & 0.77 & 0.12 & $0.03\pm_{0.02}^{0.17}$ \\
\end{tabular}
\caption{Inferred available gas fractions at the time of outflow, based on a simple spherical model, along with $\gamma_{\rm in}$ values taken from the free $f_\Upsilon$ case. Note that $M_{\rm I,gas}$ here refers to the baryon mass interior to $r_1$, so $f_{\rm g}$ is a factor 2 smaller than the total mass deficit.}
\label{fbtable}
\end{table}

The disk of DDO 154 is dominated by neutral hydrogen gas, whose presence must be accounted for when suggesting an energetic outflow \citep{carignan1998}. \cite{zubovas2011} present a simulated model of the Fermi bubbles above and below the disk of the Milky Way, detected in $\gamma$ rays \citep{su2010}. These bubbles are part of a black hole outflow that is pinched in the centre due to the density of the gas in the Galactic disk. A black hole outflow scenario is compatible with the gas-richness of DDO 154's disk if the density of that gas is high enough that an outflow able to reach $r_1$ would not significantly disturb it. If we assume that the density of the dark matter halo can be approximated as $\rho \propto r^{-1/2}$ inside $r_1$ (as an average log slope, assuming the halo becomes flat at $r=0$), and that the disk density can also be approximated near the centre of the galaxy by a $\rho \propto r^{-1/2}$ profile, the energy required to lift all matter inside $r_1$ to $r_1$ is

\begin{equation}
U = {3 \over 5} {GM(r_1)^2 \over r_1}
\end{equation}

Imparting the same amount of energy to both components, and cancelling the enclosed mass of both components, gives

\begin{equation}
{M_{\rm halo} (r_1)\over r_1} = {M_{\rm disk} (R_{\rm outflow}) \over R_{\rm outflow}} 
\label{fermieq}
\end{equation}

where $R_{\rm outflow}$ is the distance gas can be swept up in the disk with the same energy required to sweep up all halo gas to $r_1$.  For the $\rho = \rho_0 ({r \over r_0})^{-1/2}$ density profile we use, 

\begin{equation}
M = 4\pi \int_0^r \rho(r') r^2 dr = {8 \over 5}\pi\rho_0 r^{5/2} r_0^{2/3}
\end{equation}

assuming equal scale radii and substituting into (\ref{fermieq}) we can calculate how an outflow that sweeps up all gas to a specific radius would travel in the two different media

\begin{equation}
R_{\rm outflow} = \left( {\rho_{\rm 0,halo} \over \rho_{\rm 0,disk}} \right)^{2 \over 3} r_1
\end{equation}

Taking the model of \cite{banerjee2011} for the inner regions of the gas disk of DDO 154, with a pseudo-isothermal scale density of dark matter $\rho_{\rm dm} = 0.028 {\rm M_\odot  pc}^{-3}$, $\Sigma_{\rm HI} = 5.7 {\rm M_\odot pc}^{-3}$, and gas scale height $h = 100{\rm pc}$ from \cite{banerjee2011}, we derive $\rho_{\rm 0,halo} = 4.8\times10^{-3} {\rm M_\odot pc}^{-3}$ and $\rho_{\rm 0,disk} = 0.057 {\rm M_\odot pc}^{-3}$. We use the halo model of Banerjee in this calculation because their disk model is calculated assuming a pseudo-isothermal halo - however, our rotation curve data is the same. Given the value $r_1 = 1.21{\rm kpc}$ for DDO 154, we therefore estimate $R_{\rm outflow}=192{\rm {\rm pc}}$. The first radial bin in this galaxy is situated at $R = 135{\rm pc}$, which means a Fermi bubble-like outflow cannot be excluded based on the presence of gas in the disk.

\section{Conclusions}

We have applied the MCMC method described in \cite{hw2013} to a number of nearby galaxies and been able to constrain the density profiles of their haloes with less ambiguity than would be possible with simpler statistical methods. From these constraints, we have calculated the values of physical quantities ($r_1$ and $f_{\rm g}$) which can be used to constrain formation scenarios for these galaxies.  

The sample investigated here is subject to a selection bias. The THINGS galaxies were subsampled for generation of rotation curves by \cite{deblok2008}, based on inclination and other factors, and then subsampled again here on the basis of whether or not they can produce meaningful outputs from our MCMC technique. Our conclusions must be interpreted in this context.

The selection biases we experience also apply to any attempt at rotation curve decomposition. The cases where MCMC cannot find a constraint should be taken as an indication that the fitting of an individual profile, that is part of our parameter space or closely approximated by a profile that is, cannot produce a result that is credible without further discussion of the issues that prevent a constraint with MCMC. Our technique has the potential to overcome these bias using different data, or different modelling of the data (that incorporates more well constrained stellar populations for instance.) The behaviour of our MCMC technique with potential future data sets is described in HW13 and is found to be promising. 

We have identified several degeneracies in the parameter space. Some we are unable to break, such as the degeneracy between $\Upsilon$ and the inner log slope $\gamma_{\rm in}$ for NGC 3621. One of the most important degeneracies we discovered is between $\rho_{\rm s}$ and $r_{\rm s}$, which has been resolved in all cases presented here. This degeneracy precludes these two parameters being independently considered as physical. Our transformation of the parameter $\rho_{\rm s}$ into $v_{\rm max}$ removes this degeneracy, but unfortunately $r_{\rm s}$ still cannot be interpreted as a physically meaningful radius, because its position is degenerate with the shaping parameters $\alpha$, $\beta$ and $\gamma$.

Scale radii fixed by the points at which the curve reaches a particular log slope (i.e. $r_n$ where $n$ is the negative log slope) are more useful for a discussion of the actual morphology of dark matter haloes. We chose $r_1$ due to the fact that parts of the halo interior to this distance cannot be modelled by a cosmological halo such as NFW. Thus $r_1$ corresponds to a radius over which baryonic physics must act in order to produce the measured halo.

We have shown that $r_1$ is useful and well constrained, and that it can be used to constrain a simple feedback-based formation model. The relevance of $r_1$ is first as a constrainable physical parameter within the data range of the galaxies studied here, secondly as a required scale of mass loss (under reasonable assumptions) and thirdly as a common scaling parameter with which to compare observationally derived haloes to simulated ones in a physically meaningful way. The second reason applies only for $r_1$ and not for other radii. However we recognise that other radii may have similar uses, and such radii may also be constrained well with an MCMC method, as we have used here.

The model we use to derive $f_{\rm g}$ is simple, but links a physically viable outflow scenario to a quantity derived by MCMC analysis of observations, and could be refined iteratively by using it as an initial condition in formation simulations. This value cannot be reliably derived from simpler fitting methods due to the complexity of the parameter space, and the low quality of measures such as $\chi^2_{\rm red}$ as absolute goodness of fit statistics for these data. MCMC provides a firm enough constraint, and a confidence that the parameter space has been properly explored, to allow results such as $f_{\rm g}$ to guide simulations. 


\bibliographystyle{mnras}
\bibliography{citations}

\section{Acknowledgements}

This work made use of THINGS, ``The HI Nearby Galaxy Survey''~\citep{THINGS}. This work is based [in part] on observations made with the Spitzer Space Telescope, which is operated by the Jet Propulsion Laboratory, California Institute of Technology under a contract with NASA. We would like to thank Walter Dehnen, Albert Bosma and Martin Bourne for valuable discussions. The CosmoMC code was written by Anthony Lewis~\citep{lewis2002}. We would also like to thank Erwin de Blok for providing model data~\citep{deblok2008} in electronic format. MIW acknowledges the Royal Society for support through a University Research Fellowship. PRH acknowledges STFC for financial support. This research used the Complexity HPC cluster at Leicester which is part of the DiRAC2 national facility, jointly funded by STFC and the Large Facilities Capital Fund of BIS.

\end{document}